\documentclass[useAMS, usenatbib]{mn2e}

\usepackage{amsmath}
\usepackage{graphicx}
\usepackage{amssymb}
\newcommand{\comments}[1]{}
\newcommand{\dee}[0]{\mathrm{d}}
\newcommand{\new}[1]{{{#1}}}
\title[Constraining $w_{\text{DE}}$ with Compound lenses]{Constraining the dark energy equation of state with double source plane strong lenses}
\author[Collett et al.]
  {T. E. Collett$^{1}$\thanks{tcollett@ast.cam.ac.uk}, M. W. Auger$^{1}$, V. Belokurov$^{1}$, P. J. Marshall$^{2}$ and A. C. Hall$^{1,3}$\\
  $^{1}$ Institute of Astronomy, University of Cambridge, Madingley Road, Cambridge CB3 0HA\\
  $^{2}$ Department of Physics, University of Oxford, Keble Road, Oxford, OX1 3RH\\
  $^{3}$ Kavli Institute for Cosmology, University of Cambridge Madingley Road, Cambridge CB3 0HA}
\begin{document}
\pagerange{\pageref{firstpage}--\pageref{lastpage}} 

\label{firstpage}
\maketitle

\begin{abstract}

We investigate the possibility of constraining the dark energy equation of state by measuring the ratio of Einstein radii in a strong gravitational lens system with two source planes. This quantity is independent of the Hubble parameter and directly measures the growth of angular diameter distances as a function of redshift. We investigate the prospects for a single double source plane system and for a forecast population of systems discovered by re-observing a population of single source lenses already known from a photometrically selected catalogue such as {\sc CASSOWARY} or from a spectroscopically selected catalogue such as SLACS. We find that constraints comparable to current data-sets ($\sigma(w)\sim15\%$) are possible with a handful of double source plane systems. We also find that the method's degeneracy between $\Omega_{\text{M}}$ and $w$ is almost orthogonal to that of CMB and BAO measurements, making this method highly complimentary to current probes.
 
\end{abstract}


\section{Introduction}

We appear to live in an expanding universe, and the expansion is accelerating \citep{riess, perlmutter}. This acceleration cannot be caused by the baryonic matter that we interact with everyday, because it exerts a braking effect on the universe's expansion; instead, the acceleration must be caused by something (frequently termed dark energy) that exerts a negative pressure on the universe. Our current understanding of dark energy is very poor; from observations we know that it today makes up around 70\% of our universe's energy density \citep{larson} and has an equation of state with $w \approx -1$, where $w$ is defined as the ratio of pressure to energy density, $w~=~p/{\rho c^2}$. Any equation of state with $w \le -{1 \over 3}$ will exert sufficient negative pressure to cause accelerated expansion, so why do we currently observe $w \approx -1$? 

Little is known about dark energy so there are many possible explanations that we cannot currently exclude. The standard model of a cosmological constant (with $w~=~-1$) is consistent with the current best constraints \citep{eisenstein, percival, komatsu, suyu}, but the current uncertainties leave much room for manoeuvre, including models where the equation of state varies with time \citep{caldwell}. Tighter constraints will allow us to better pin down the nature of dark energy and the role it plays in the expansion of our universe. 

\begin{figure}
\includegraphics[width=\columnwidth]{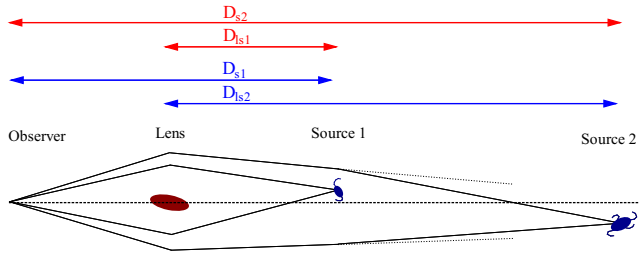}
\caption {Sketch of a double source plane lens system. For a singular isothermal sphere, the dimensionless number $\eta$ is the product of $D_{\text{ls}1}$ and $D_{\text{s}2}$ (both in red) divided by the product of $D_{\text{ls}2}$ and $D_{\text{s}1}$ (both in blue).}
\label{bench}
\end{figure}

First discovered in Q0957+561 by \citet{walsh}, strong gravitational lensing is now a powerful cosmological tool \citep[e.g.][]{witt, kochanek, saha, schechter, oguri, congdon2008, congdon2010, keeton, suyu} and the time delay strong lensing constraints on dark energy are forecast to improve \citep{coe}. In this this work we examine how well $w$ can be constrained using a population of gravitational lens systems with two background sources at different redshifts (schematic shown in Figure~\ref{bench}). Such systems are known with cluster scale lenses \citep[e.g.,][]{soucail, jullo} and a galaxy--scale double source plane system was recently found \citep{gavazzi}, although cosmography is currently imprecise using the galaxy--scale system, as no spectroscopic redshift for the far source has been obtained \citep[but see][]{sonnenfeld}. The incentive to use such systems is strong; current methods of constraining $w$ are degenerate with the Hubble constant \citep{eisenstein, percival, komatsu, suyu,weinberg}, but taking the ratio of the two Einstein radii in a double source system allows us to constrain $w$ independently of the Hubble constant. This independent constraint will not only teach us about the nature of dark energy, but will also help improve constraints on other cosmological parameters by providing complimentary probes with a prior on $w$. For example the WMAP7 constraints on $h$ (the reduced Hubble constant) are $h~=~0.710 \pm {0.025}$ \citep{komatsu} for a flat $\Lambda$CDM model with $w$ fixed at $-1$, whilst allowing $w$ to be a free parameter loosens the WMAP-only constraints by nearly an order of magnitude to $h~=~0.75_{-0.14}^{+0.15}$.

This paper is meant to investigate the cosmological information content of double source plane lenses; we intend to address the following questions:

\begin{enumerate}
\item {\it Can double source plane lenses be used for cosmography?}
\item {\it What is an optimal configuration of lens and source redshifts for cosmography?}
\item {\it How well could cosmography be constrained with a realistic population of double source plane lenses?}\\
\end{enumerate}

\noindent In future papers we will investigate the observational difficulties and potential systematics, as well as producing detailed forecasts for a double source plane search with upcoming facilities.

The paper is laid out as follows. In Section 2 we outline the method and relevant quantities. Section 3 looks at the prospects for cosmography with a single hypothetical lensed double source system and should provide the reader with an intuition of what is an optimal system for cosmography, whilst Section 4 investigates cosmography with a population of double source plane lenses. The population of double sources are forecast for re-observing a pre-existing lens-source catalogue. In Section \ref{complications} we examine the cosmographic constraints for an evolving equation of state. Section 6 concludes by examining the effectiveness of combining a set of double source plane systems with current constraints upon $w$ from other probes.

Throughout this work we assume a spatially flat fiducial cosmology with $w~=~-1$ at all times and $\Omega_\text{M}$~=~0.27; where necessary we take $h~=~0.7$. All models assert flatness and have a constant equation of state, except in Section \ref{complications}, where we investigate an evolving model and spatial curvature.

\section{Gravitational lensing and double source systems}\label{onesys}

Mass causes the deflection of light and so can act as a gravitational lens, producing a set of images where the time delay function is stationary. For a strong lens system with a source lying directly on the optical axis this produces an Einstein ring. The Einstein radius is sensitive to the angular diameter distances between observer, lens and source. In the case of a thin lens in an otherwise homogeneous universe the Einstein radius is given by
\begin{equation}
{\theta_{\text{E}}~=~\sqrt{{4GM(\theta_{\text{E}})\over{c^2}}{{D_{\text{ls}}} \over {D_{\text{ol}}D_{\text{os}}}}}}
\label{einstein radius}
\end{equation}
where $\theta_{\text{E}}$ is the Einstein radius, ${M}(\theta_{\text{E}})$ is the projected mass within the Einstein radius, and the $D_{ij}$ are the angular diameter distances between observer (o),\footnote[1]{From now on we drop the `o' referring to the observer in observer-lens/source distances} lens (l) and source (s). In this work we assume all lenses to be Singular Isothermal Spheres (SISs), with a spherically symmetric mass distribution. \new{The density of an SIS profile is given by:}
\begin{equation}
\rho(r)~=~{{\sigma_V}^2 \over {2 \pi G r^2}}\label{SIS}.
\end{equation}
SIS mass distributions have been shown to approximate the observational data well \citep[e.g.][]{koopmans, auger2010} and allow for simple lensing forecasts since the lensing deflection angle and magnification have analytical forms. The Einstein radius of an SIS is given by
\begin{equation}
\theta_{\text{E}}^{\text{SIS}}~=~{4 \pi{{\sigma_V^2}\over{c^2}}{{D_{\text{ls}}} \over {D_{\text{s}}}}}~\simeq~{\left({{\sigma_V}\over{186~\text{km s} ^{-1}}}\right)^2 {{D_{\text{ls}}} \over {D_{\text{s}}}}} \text{ arcseconds.}
\label{SISeinstein radius}
\end{equation}
For an SIS model, strong lensing occurs when the source is in the region $0<\theta_{\text{s}}^{\text{}}<\theta_{\text{E}}$, where $\theta_{\text{s}}^{\text{}}$ is the {\it unlensed} angular position of the source with respect to the optical axis. Two images are produced at $\theta_{+, -}~=~ \theta_{\text{s}}^{\text{}} \pm \theta_{\text{E}}$ with magnification $\mu_{+, -}~=~ 1 \pm \theta_{\text{E}}/\theta_{\text{s}}^{\text{}}$.

Since the equation of state governs the expansion of our universe it affects the evolution of angular diameter distances as a function of redshift. 
\begin{equation}
D_{\text{ij}}~=~ {c/H_0 \over {(1+z_{\text{j}})}}\left( {\mathrm{sinn}\!\left( \sqrt{|\Omega_k|} \int_{z_{\text{i}}}^{z_{\text{j}}} {\mathrm{d} z \over E(z)}\right)\over \sqrt{|\Omega_k|}}\right)
\label{Da}
\end{equation}
where $\mathrm{sinn}(x)~=~\sin(x)$, $x$, or $\sinh(x)$ for open, flat, or closed universes respectively, and $E(z)$ is the normalised Hubble parameter:
\begin{subequations}
\begin{align}
E(z) &\equiv {H(z) \over H_0}\\
E^{w\text{CDM}}&~=~\!\sqrt{{\Omega_{\mathrm{M}}(1\!+\!z)^{3}\!+\!\Omega_k(1\!+\!z)^2\!+\!(\Omega_{\mathrm{de}})(1\!+\!z)^{3(1+w)}}}\label{Ez}\\
E^{w_z\text{CDM}}&~=~\!\sqrt{{\Omega_{\mathrm{M}}(1+z)^{3}+\Omega_k(1+z)^2+(\Omega_{\mathrm{de}})e^{I(z)}}}.
\label{Ez2}
\end{align}
\end{subequations}
$I(z)$ is an integral given by
\begin{equation}
I(z)\equiv 3\int_0^z{1+w(z') \over(1+z')} \dee z'.
\label{Iz}
\end{equation}
Equation \ref{Ez} holds if $w$ is constant, whilst Equation \ref{Ez2} is general for any universe with a time-evolving equation of state. Neglecting the mass of the closer source, we define the quantity $\eta$ as the ratio of the two Einstein radii,
\begin{equation}
\eta~=~{\theta_{\text{E}, 1} \over \theta_{\text{E}, 2}} 
\end{equation}
with s1 and s2 referring to the near and far source, respectively. 
For an SIS lens $\eta$ is given by
\begin{equation}
\eta^{\text{SIS}}~=~{{{D_{\text{ls}1}D_{\text{s}2}}\over{D_{\text{ls}2}D_{\text{s}1}}}}.
\end{equation}
The ratio, $\eta$, has the intrinsic advantage that it is independent of the Hubble constant and is only weakly dependent upon the mass distribution of the lens (in the case of an SIS model $\eta$ is independent of the mass), so is a function only of $w$, $\Omega_\text{M}$ and the redshifts of the lens and sources (and the mass model).
\comments{
\begin{figure}
\includegraphics[width=\columnwidth]{Rw.eps}
\caption {$\eta$ as a function of $w$, in an otherwise fixed universe (black). For reference the blue, red and green lines show how $\eta(w)$ varies with decreasing $z_l$, decreasing $z_{\text{s}1}$ and increasing $z_{\text{s}2}$ respectively. Systems where the gradient of $\eta(w) /  \eta(w~=~-1)$ is steepest about $w~=~-1$, will be most useful for cosmography, as these represent systems where $w \neq -1$ significantly change $\eta$.}
\label{etavsw}
\end{figure}
}

{
\begin{figure}
  \begin{center}
    \begin{tabular}{c}
      \resizebox{85mm}{!}{\includegraphics{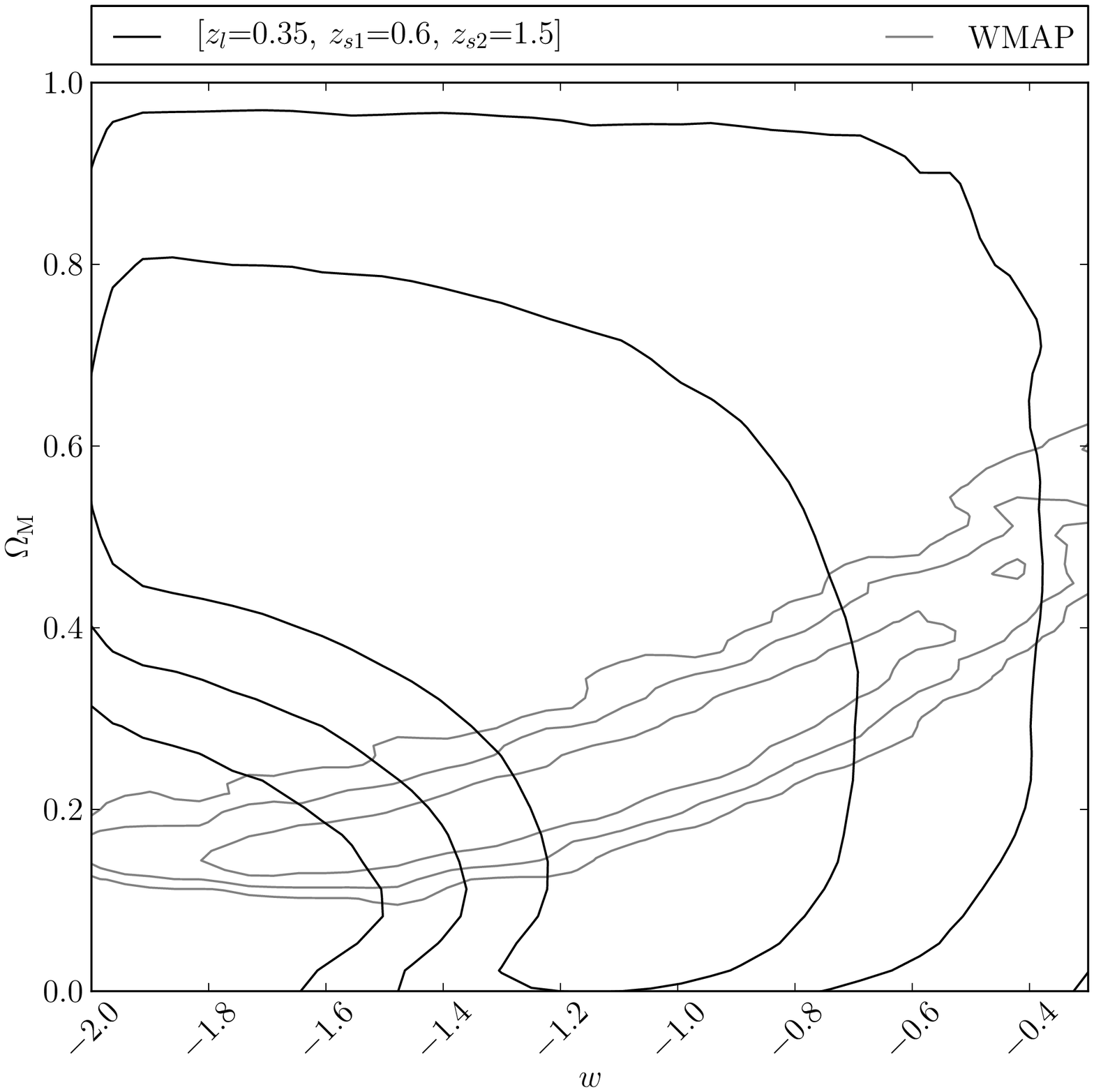}}\\
    \end{tabular}
  \end{center}
\caption{The $w$ and $\Omega_{\text{M}}$ plane: black are the constraints possible from our fiducial single double source system ($z_{\text{l}}~=~ 0.35, z_{\text{s}1}~=~0.6, z_{\text{s}2}~=~1.5$), given uniform priors of $0 \le \Omega_{\text{M}} \le 1$ and $-2 \le w \le -{1 \over 3}$ and a measurement with $\sigma(\eta)/(\eta)~=~1\%$. Grey shows the WMAP7 results. Contours are 1, 2 and 3 $\sigma$. The orthogonality (and hence complementarity) between WMAP and the double source plane constraint is striking.
}\label{sim}
\end{figure}
}

\section{Single system constraints on a constant equation of state}

In order to investigate $[w$, $\Omega_\text{M}]$ we first fix our lens and source redshifts to gain intuition about the problem. We choose a hypothetical double source plane lens, with [$z_{\text{l}}~=~ 0.35, z_{\text{s}1}~=~0.6, z_{\text{s}2}~=~1.5$] as the ``fiducial system". We use a uniform prior of $0 \le \Omega_{\text{M}} \le 1$ and $-2 \le w \le -{1 \over 3}$ and assume the observational precision on $\eta$ to be 1\% (i.e. $\sigma(\eta)/(\eta)~=~0.01$) to perform a Markov Chain Monte Carlo (MCMC) analysis of the $w$--$\Omega_\text{M}$ plane. \new{For our MCMC analysis we assume Gaussian likelihoods given by 
\begin{equation} L~\propto~\exp \left[ -{1 \over 2}{{(\eta(\Omega_{\text{M}}, w)-\eta_{\text{Fiducial}})^2}\over{\sigma^2}}\right] \end{equation}\label{Li}

\noindent where $\eta_{\text{Fiducial}}$ is the value of $\eta$ expected for the fiducial cosmology.}
The resulting 1, 2, and 3 $\sigma$ contours are shown in Figure~\ref{sim}. Given the assumed 1\% measurement, we find that it would be possible to constrain $w < -0.54$ at the 95\% CL, for the fiducial system. However the posterior gives little constraint on $w\lesssim-0.7$ without a tighter prior on $\Omega_\text{M}$; this occurs because in a high $\Omega_\text{M}$ universe, dark energy plays very little role in the expansion history and so $\eta$ becomes broadly insensitive to the equation of state. A significant body of evidence now exists to suggest $\Omega_{\text{M}} \sim 0.3 $ \citep[e.g.,][]{larson}; this allows us to exclude the right of Figure~\ref{sim} and places our universe in the regime where $\eta$ {\it is} sensitive to the dark energy equation of state.
{
\begin{figure}
\includegraphics[width=\columnwidth]{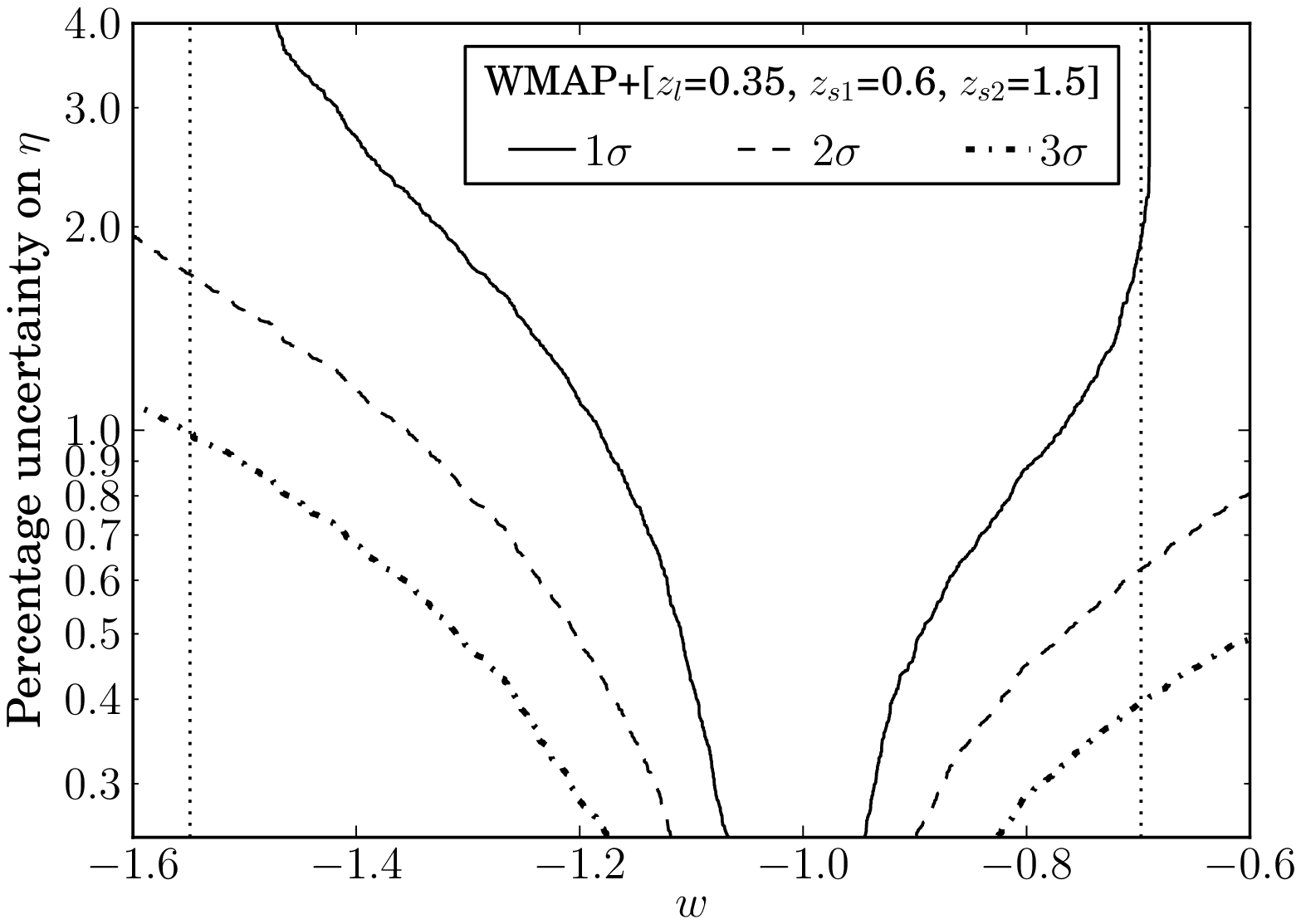}
\caption{The forecast 1, 2 and 3 $\sigma$ confidence region on $w$ possible from a single double source system ($z_{\text{l}}~=~ 0.35, z_{\text{s}1}~=~0.6, z_{\text{s}2}~=~1.5$), as a function of the uncertainty on $\eta$, marginalised over all $\Omega_{\text{M}}$ using a WMAP prior. The dotted vertical lines represent the 1$\sigma$ constraint on $w$ from WMAP alone. \new{Uncertainties significantly below 1\% are currently unrealistic given the need to correctly model the compound lensing introduced by the first source. The inferred uncertainty on $\eta$ in \citet{gavazzi} is 1.2\%, without accounting for lensing by the intermediate source but only using a conjugate points analysis rather than all available pixels from the HST images.}}\label{unc}
\end{figure}
}
In order to take advantage of this prior knowledge we use the WMAP 7 year results from \citet{komatsu} and resample their MCMC chains\footnote[2]{lambda.gsfc.nasa.gov/product/map/dr4/parameters.cfm} using the importance sampling method as outlined in \citet{lewis}. We use the MCMC chains produced assuming a flat $w$CDM model, giving a set of points each possessing a value for  $\Omega_{\text{M}}$, $w$ and a weight. \new{Each point in the WMAP chain is then given a likelihood $L_i$, according to Equation~\ref{Li} and the chain is re-weighted such that the posterior is the product of the original weight and the likelihood.} In Figure~\ref{unc} we plot the 1, 2, and 3 $\sigma$ confidence intervals for the fiducial system plus WMAP as a function of percentage uncertainty on $\eta$; an uncertainty of less than 2\% is key to improving the constraint on $w$ with a single system. Whilst minimising the uncertainty on $\eta$ maximises the constraint on $w$, a measurement much more accurate than 1\% is unlikely with current resources, given the need for space telescope time in order to constrain well the lens mass model and the image separation. The inferred uncertainty on $\eta$ in \citet{gavazzi} is 1.2\%, neglecting any uncertainties induced by lensing from the first source. These systems are compound lenses;
\citet{gavazzi} use the Tully-Fisher relation to estimate the mass of the first source and
conservatively estimate that the uncertainty on $\eta$ grows to 
approximately 6 per cent. Their estimate does not attempt to model the relensing and hence neglects most of the information available in the high-resolution imaging, and modeling the lensing by the first source of a typical compound lens is likely to significantly improve upon the 6\% estimate of \citet{gavazzi}. \new{It has been shown \citep[e.g.][]{vegetti} that using a pixelated image modelling technique can quantify perturbations due to substructure with masses as low as 0.1\% of the the primary lens mass (enclosed within the Einstein radius), although this method has not yet been applied to significant perturbers outside the primary lens plane}. In this work we take $\sigma(\eta)/(\eta)~=~1\%$ unless otherwise stated; this may prove optimistic, but serves to illustrate the cosmological information encoded in such systems. We do not wish to imply that 1\% is definitely achievable and we will explore the precision with which $\eta$ can be measured in future work \new{(Collett et al. In prep)}.
\begin{figure*}
  \begin{center}
      \resizebox{55mm}{!}{\includegraphics{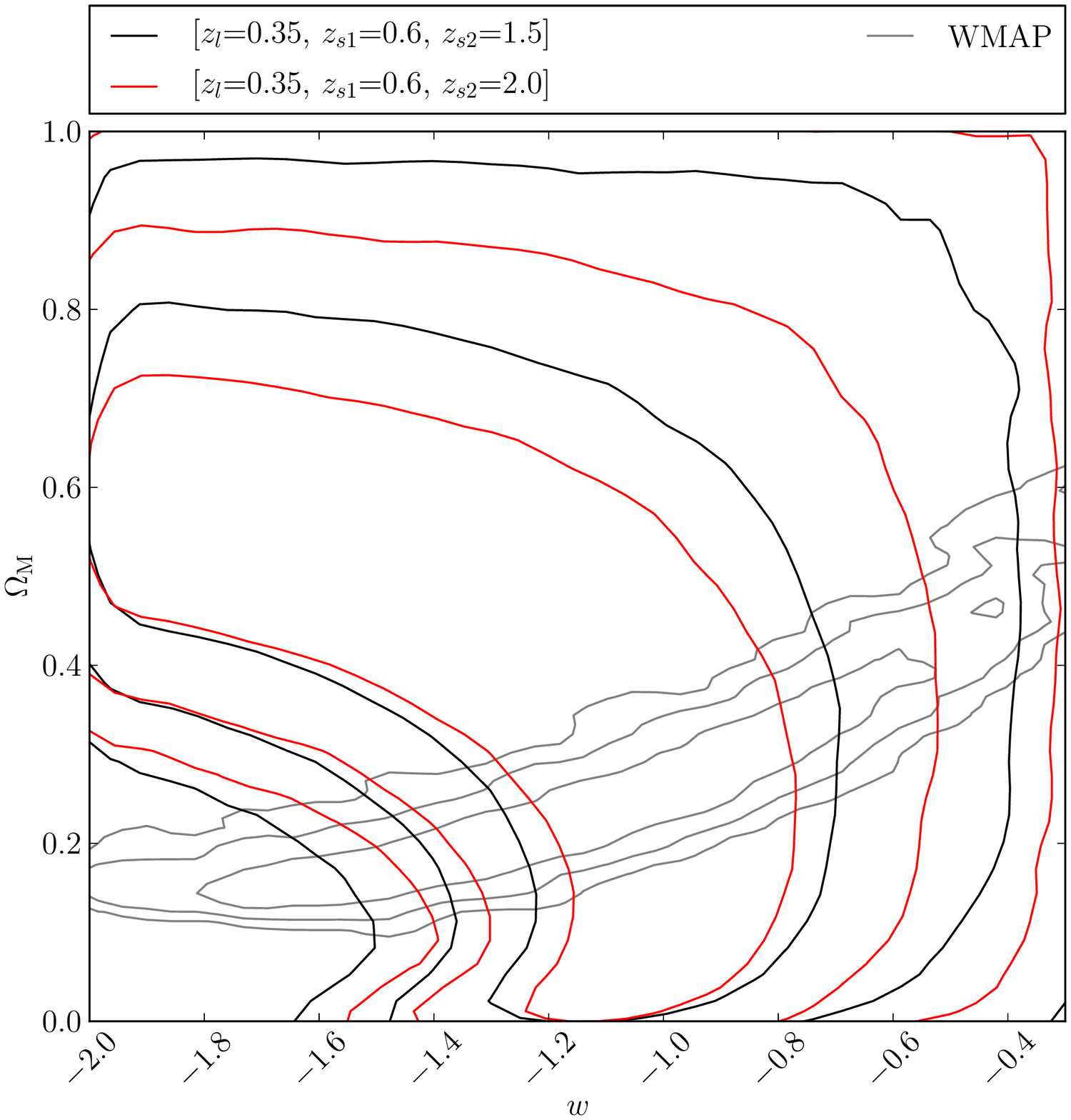}} 
      \resizebox{55mm}{!}{\includegraphics{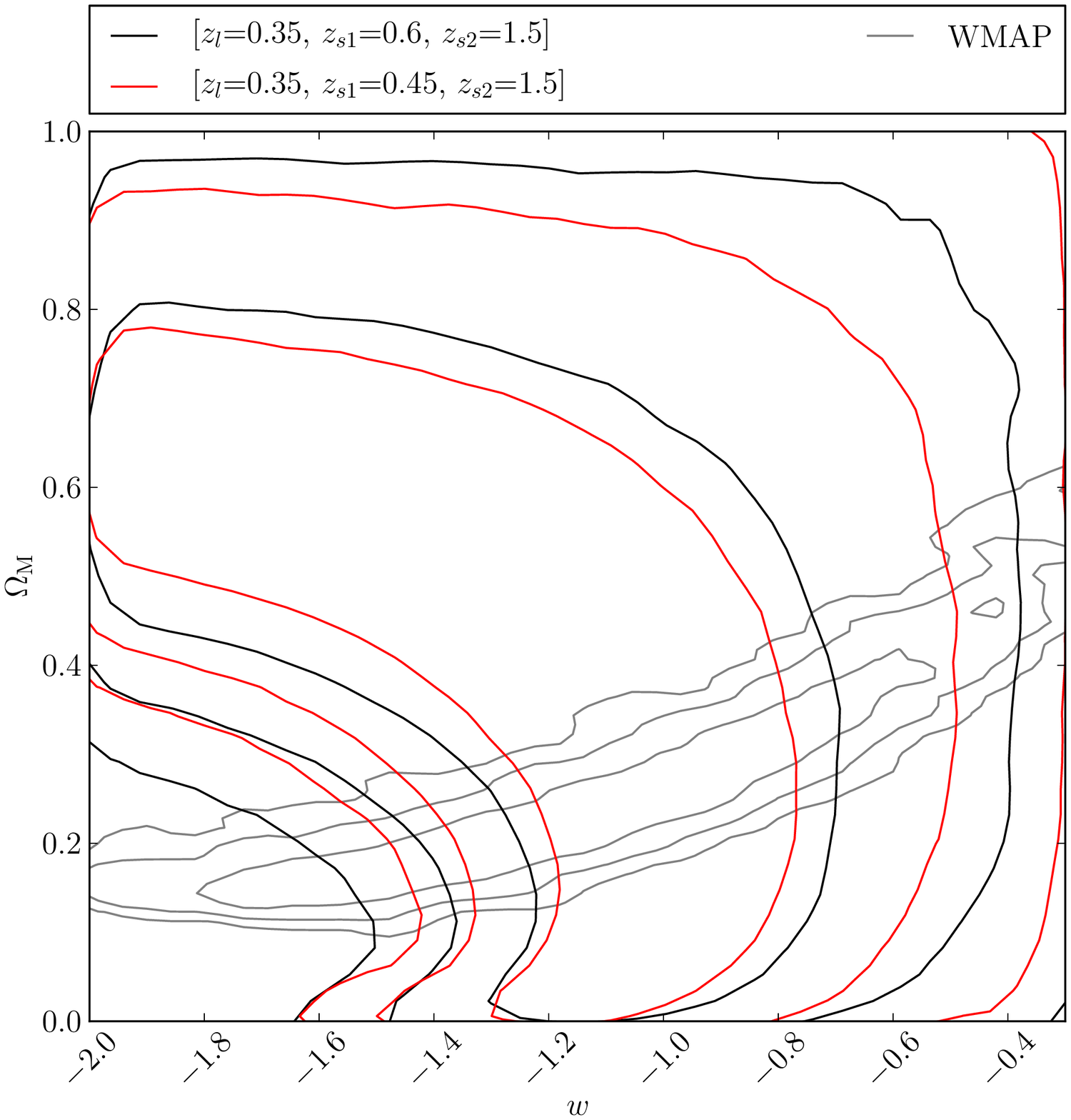}} 
      \resizebox{55mm}{!}{\includegraphics{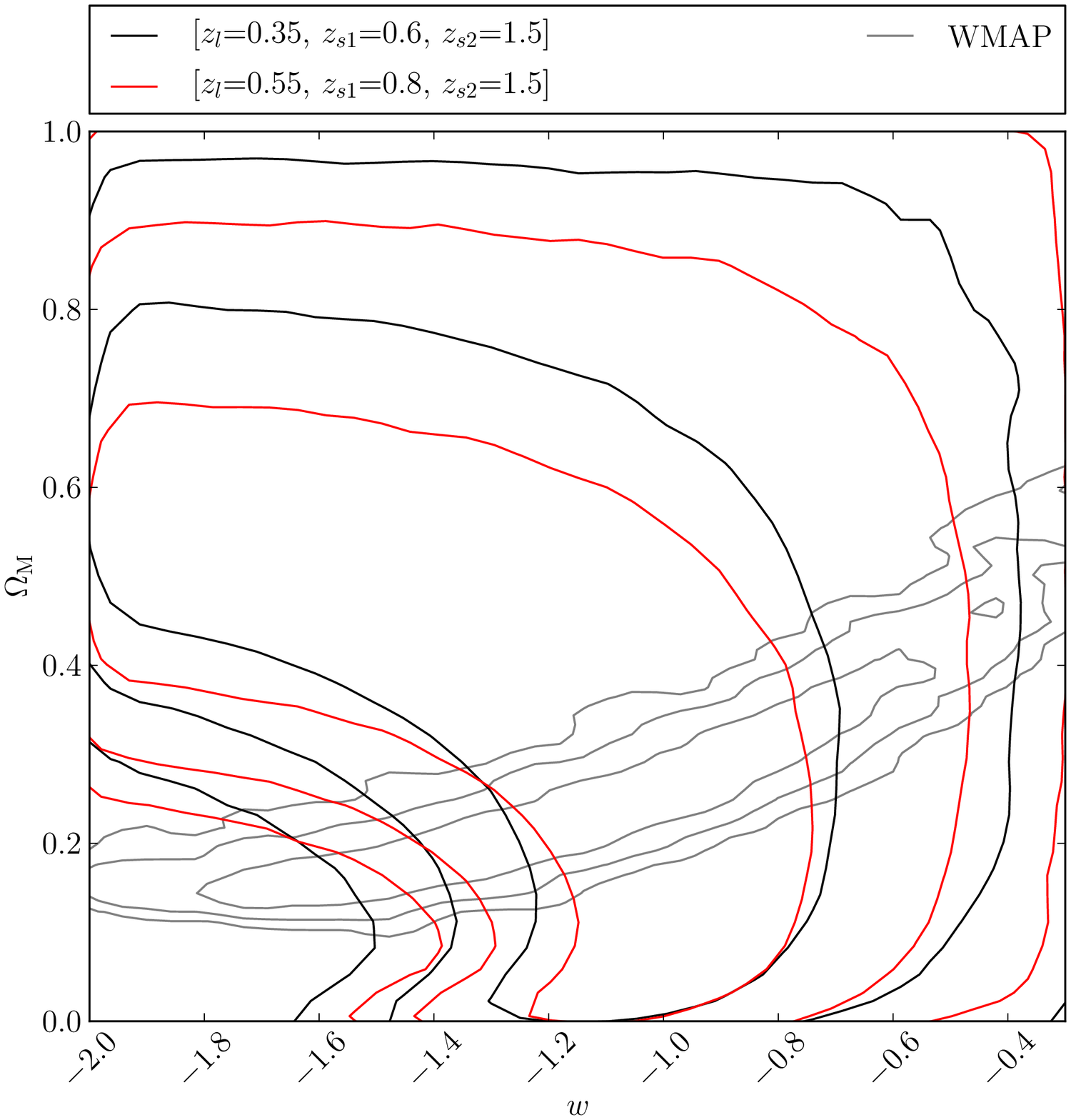}}
  \end{center}
\caption{Same as Figure~\ref{sim} but with a different double source plane system over-plotted in red. {\it Left:} The second source moved further away \new{($z_{\text{l}}~=~0.35$, $z_{\text{s}1}~=~0.6$, $z_{\text{s}2}~=~2.0$)}. {\it Centre:} The first source moved closer to the lens  \new{($z_{\text{l}}~=~0.35$, $z_{\text{s}1}~=~0.45$, $z_{\text{s}2}~=~1.5$)}. {\it Right:} The lens and first source moved to a redshift of 0.2 greater  \new{($z_{\text{l}}~=~0.55$, $z_{\text{s}1}~=~0.8$, $z_{\text{s}2}~=~1.5$)}. One can infer from these that the optimal system has a close pair of lens and first source, with a distant background source. Whilst moving the lens and first source to higher redshift alters the tilt of the banana, there is only marginal improvement on the constraints in the region that is also consistent with WMAP7.}\label{move}
\end{figure*}
{
\begin{figure*}
  \begin{center}
      \resizebox{75mm}{!}{\includegraphics{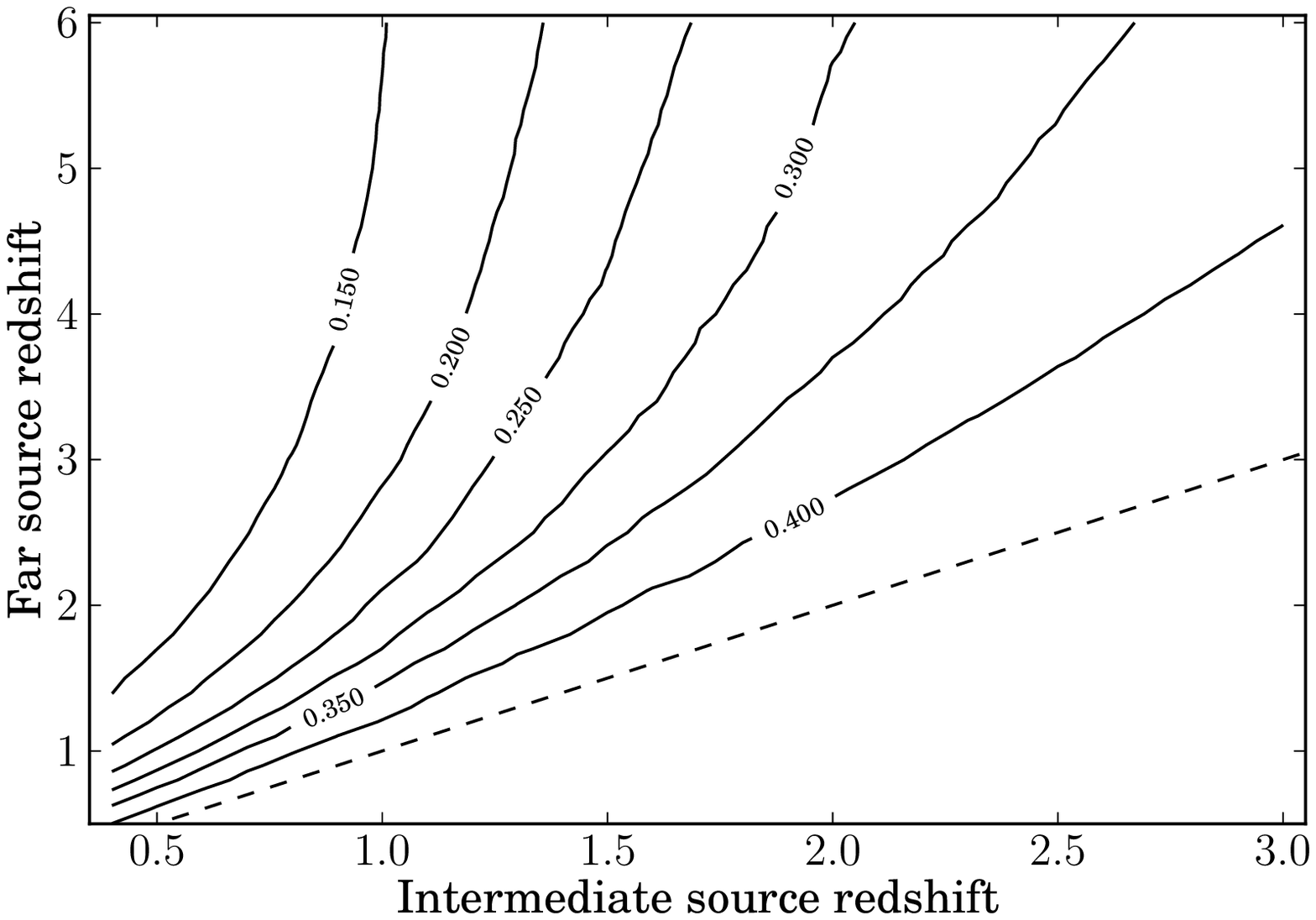}}
      \hspace{10 mm}
      \resizebox{75mm}{!}{\includegraphics{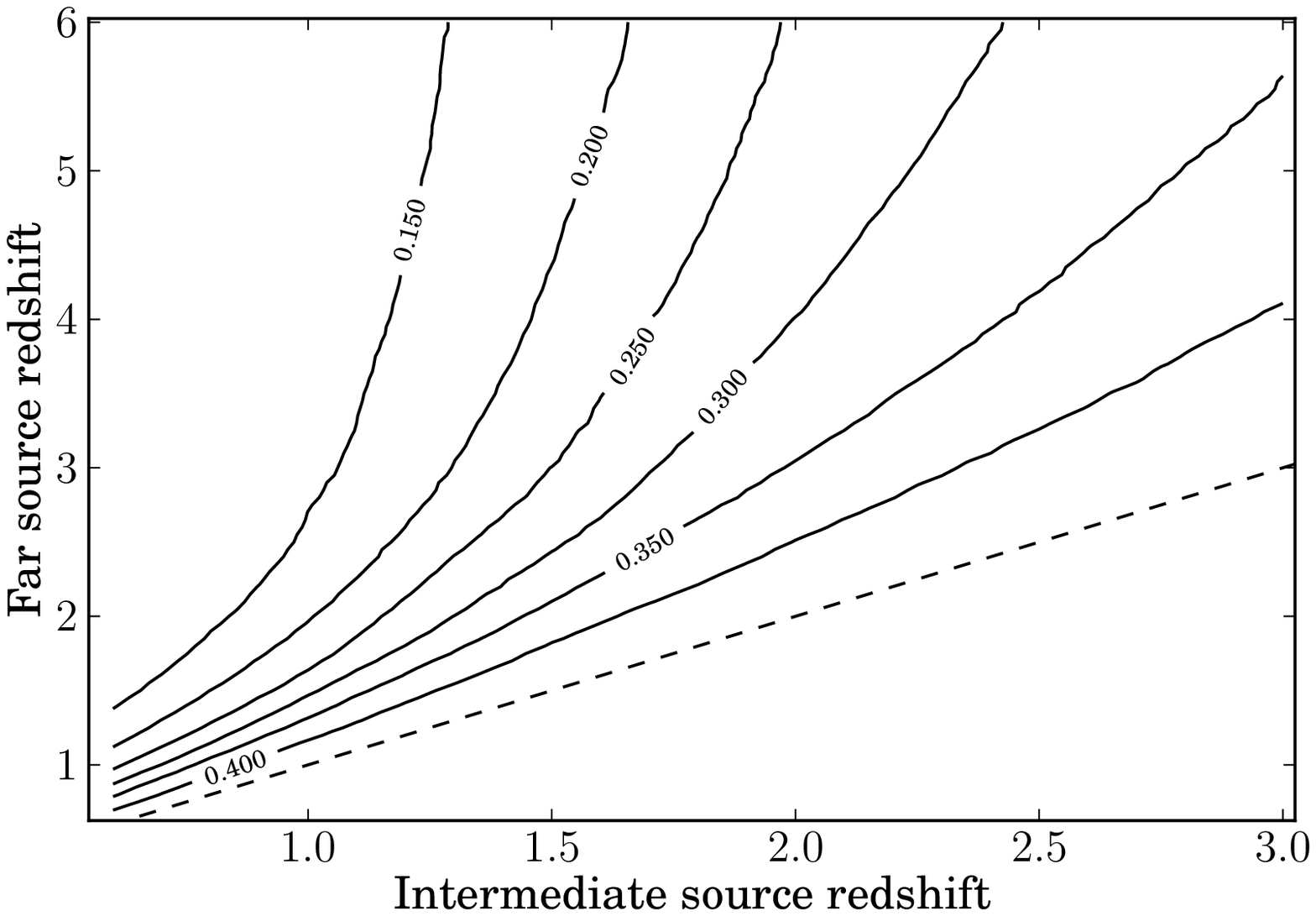}}
  \end{center}
\caption{Contour plots of the symmeterised 1$\sigma$ constraints on $w$ possible for a single double source plane system combined with WMAP, for fixed lens position and variable source positions, given an uncertainty of $\sigma(\eta)/\eta~=~1\%$. The dashed line indicates $z_{\text{s1}}~=~z_{\text{s2}}$. {\it Left:} Lens redshift $z_{\text{l}}$~=~0.35. {\it Right} Lens redshift $z_{\text{l}}$~=~0.55. The contours show that maximising the separation between first and second source and minimising the separation between lens and first source gives the best constraints on $w$.}\label{intuition}
\end{figure*}
}
A measurement of the fiducial system at $\sigma(\eta)/(\eta)~=~1\%$ gives the constraint of $w~=~-0.99_{-0.22}^{+0.19}$ when combined with the WMAP7 dataset. The main improvement upon the WMAP data comes from the fact that low $w$, low $\Omega_{\text{M}}$ universes (bottom left of Figure~\ref{sim}) are strongly excluded by the double source plane system given our fiducial cosmology.

In order to further build intuition and to suggest an observing strategy, we investigate the optimal [$z_{\text{l}}, z_{\text{s}1}, z_{\text{s}2}$] configuration for a single system. Figures~\ref{move} and ~\ref{intuition} show the effect of altering the redshifts of each object in the system. We find that increasing the redshift of the second source improves the constraint on $w$; this makes physical sense since it means $D_{\text{s2}}$ and $D_{\text{ls2}}$ probe a longer part of the cosmic expansion history, amplifying any deviations from $D_{\text{s2}}^{w=-1}$. We also find that the constraint on $w$ is improved by decreasing the separation of lens and first source. Small $\Delta z_{\text{l-s1}}$ and large $\Delta z_{\text{s1-s2}}$ being the optimal configuration is physically sensible -- it makes ${D_{\text{ls}2}} / {D_{\text{ls}1}}$ most sensitive to $w$, since very little of the ${D_{\text{ls}2}}$ integral is spent in the redshift range also included in ${D_{\text{ls}1}}$. It is also important that $z_{\text{l}} \gtrsim 0.2$ or the ratio of $D_{\text{s}} / D_{\text{ls}}$ quickly saturates to $1$ independent of the underlying cosmography, ensuring $\eta \approx 1$; this is also the case when $\Delta z_{\text{s1-s2}}$ is small.

Small $z_{\text{s}1}-z_{\text{l}}$ also increases the importance of lensing by the intermediate object and could introduce a significant systematic error if the lensing by the first source is modeled incorrectly. We leave the analysis of secondary lensing for a later work, but note that it should be possible to model for bright systems due to the wealth of positional information contained in two bright arcs. For low mass first sources secondary lensing is likely to be dominated by external convergence between the sources, which can be statistically modelled \citep{suyu}. Secondary lensing is likely to cause problems in systems where the intermediate source is of intermediate mass and the background source is faint. A second factor that may trouble systems with small $z_{\text{s}1}-z_{\text{l}}$ is that the mass density profile within the inner Einstein ring may be significantly different to that contained by the outer ring and this will mean the uncertainty on $\eta$ could be significantly affected by the uncertainty on the mass distribution of the lens. Hence, for very small $\Delta z_{\text{l}-\text{s1}}$ it may be impossible to achieve a 1\% uncertainty on $\eta$, although the cross section for very small $\Delta z_{\text{l-s1}}$ is also very small.

\section{Prospect for cosmography with n double source systems}
Angular diameter distance is an integral of $w$ and $\Omega_{\rm M}$ over redshift, so the exact nature of the degeneracy between the two parameters varies with $z_{\text{l}}, z_{\text{s}1}, $ and $z_{\text{s}2}$. The curve of the ``banana" shaped distribution in Figure~\ref{sim} depends on the redshifts, as does the constraining power of each system; hence we do not necessarily expect $n$ systems to provide a $\sqrt{n}$ improvement. Indeed, the precision could scale better or worse than $\sqrt{n}$ depending on the widths and orientations of the constraints from each system; the constraining power of $n$ systems is not trivial to predict. 

We now draw a set of double source systems from a realistic population of lenses and sources. We assume that double lenses are more likely to be found by observing already-known single lens systems to greater depth and at different wavelengths. We focus on two types of initial surveys: photometrically selected wide arc catalogues where the survey looks for multiple sources around a central object \citep[e.g., {\sc CASSOWARY};][]{belokurov2007} and spectroscopically selected catalogues where the survey looks at the spectra of galaxies for higher redshift emission lines, such as SLACS \citep{bolton2004}.

Follow up of an existing catalogue introduces two selection functions, the selection function of the original survey and of the follow up. We estimate the selection function for current surveys in two ways: we create a mock photometric catalogue for {\sc CASSOWARY}-like surveys and we draw random systems from the current SLACS sample of \cite{auger2009}. For the follow up selection function we assume the observing strategy will prioritise those systems which are best for cosmography. In practice this usually means looking for high-redshift background second sources given a low redshift ($z_{\text{l}}, z_{\text{s}1}<1.5$) lens--source pair, or low redshift intermediate sources, given a well separated lens and source. As such we assume follow up will either be conducted in the millimetre, in order to preferentially select higher redshift second sources, or in deep optical imaging, where low-redshift intermediate sources could be found. For the millimetre population we adopt the unlensed model of \citet{bethermin}, assuming observations at 1.1~mm and continuum flux limits of 0.3~mJy and 1~mJy; an example source redshift probability distribution function for a lens at $z_{\text{l}}~=~0.4$ is shown in Figure~\ref{bethpop}.{
\begin{figure}
\includegraphics[width=\columnwidth]{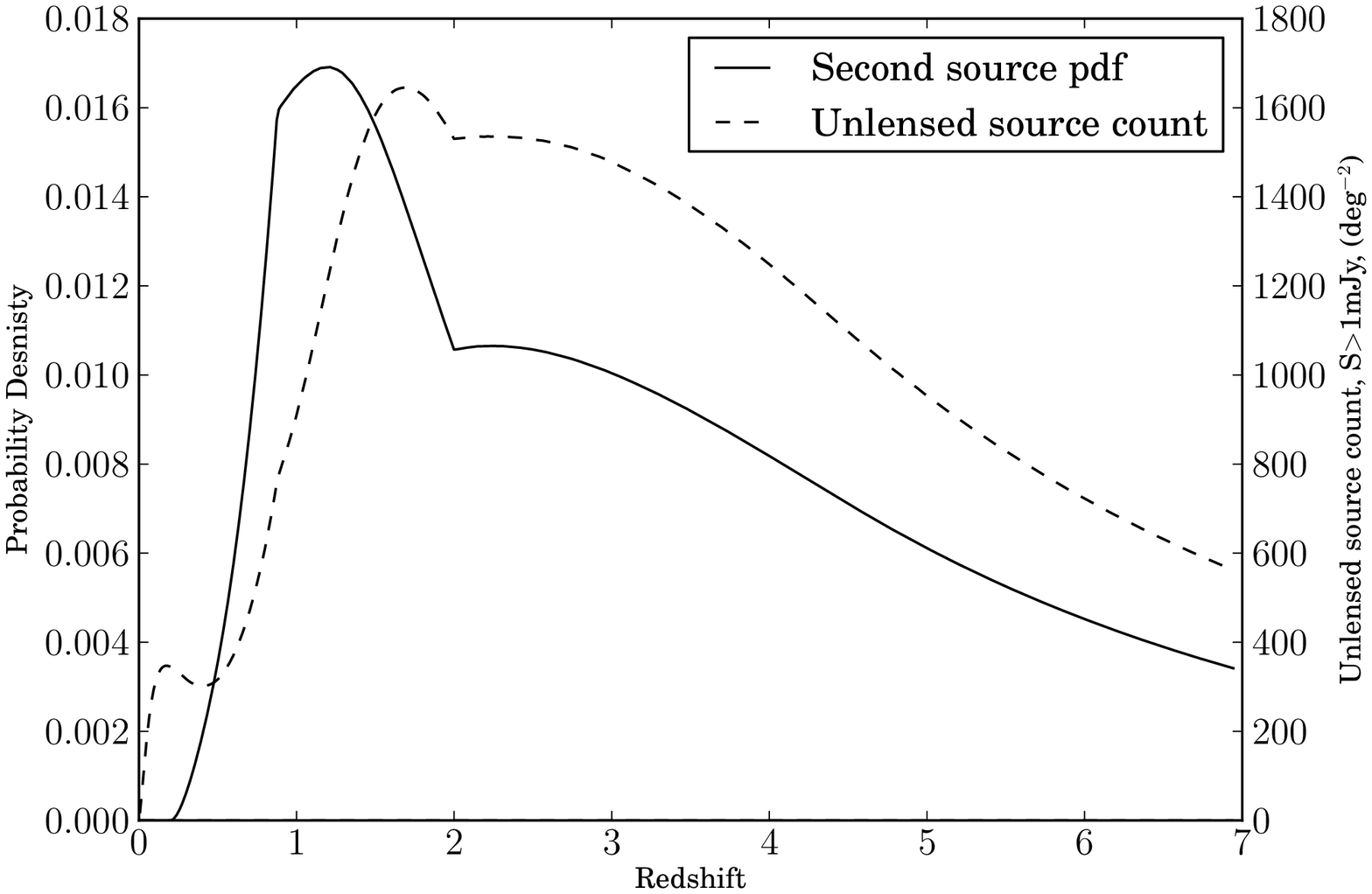}   
\caption{Example probability distribution function for finding a second source behind a lens at $z_{\text{l}}~=~0.2$, with velocity dispersion 350~km~s$^{-1}$. given observations at 1.1~mm down to a flux limit 0.3~mJy. The PDF does not integrate to one, since it is not certain that a second source is found. The dashed line shows the unlensed number density of sources with $S~>~0.3$~mJy per square degree (scale on right hand $y$-axis) in the \citet{bethermin} model. Some of the $S~<~0.3$~mJy objects also contribute to the lensed PDF due to magnification.} \label{bethpop}
\end{figure}
} For the optical source distribution we use the photometric redshift catalogue of galaxies observed in the COSMOS \citep{cosmos} field; this catalogue goes to an $i$-band depth of 25 mag, but we limit our detection depth to an $i$ magnitude of 22 to allow for magnifications of $\sim 15$ by strong lensing. Both the optical and millimetre depths are conservative in light of upcoming telescopes. However, these represent the limit to which we trust the models; beyond these depths a currently unknown population of dim objects may exist, although if the shape of the redshift distributions do not vary significantly with flux limit then the cosmological conclusions are independent of survey depth. Given a population of known lenses we evaluate the probability of being a double source plane lens, $P(D|L)$: for a $[z_\text{l}$, $\sigma_{V}]$ pair we calculate the lensing cross section and magnification tensor as functions of redshift, and apply these to the \citet{bethermin} model or COSMOS catalogue. This method is broadly similar to the method in \citet{OandM}. For axi-symmetric lenses our method is equivalent to solving for each lens the integral
\begin{equation}
P(D|L)~=~ \int _{z_{\text{l}}} ^{\infty} \dee z {\int _{0} ^{\infty} \dee S \, \sigma_{\text{cross}}(z, z_{\text{l}}, \sigma_V) {{\dee^2N}\over{\dee z \dee S}}W(S, S_{\text{lim}}) }
\label{integral}
\end{equation}
where ${\dee^2N}\over{\dee z \dee S}$ is the differential number distribution of second sources per unit solid angle with respect to redshift and flux, $\sigma_{\text{cross}}$ is the lensing angular cross section, and $W$ is a function of the flux ($S$) and the survey flux limit ($S_{\text{lim}}$); $W$ encodes the magnification bias. In the case of an SIS,
\begin{equation}
\sigma_{\text{cross}}(z, z_{\text{l}}, \sigma_V)~=~\pi \left[\theta_{\text{E}}^{\text{SIS}}(z, z_{\text{l}, \sigma_V)}\right]^2
\end{equation}
and
\begin{equation}
W(S, S_{\text{lim}})~=~ 
  \begin{cases}
   1  & \text{if } S \geq S_{\text{lim}}/2 \\
   \left({2S\over S_{\text{lim}}}\right)^2      & \text{if } S < S_{\text{lim}}/2
  \end{cases}
\label{W(S)}
\end{equation}
with Equation \ref{integral} gives the probability for finding a second source behind each lens. The probability distribution function (PDF) for finding a source at redshift $z_{\text{s2}}$ is the derivative of Equation \ref{integral} with respect to $z$. We then generate the second source redshift by drawing uniformly from the PDFs to generate a population of [$z_{\text{l}}, z_{\text{s}1}, z_{\text{s}2}$] systems, with a weight given by $P_i(D|L)$. Given the weighted population, we sample 1000 sets of $n$ systems from the population, and forecast the limits on $w$ for each set. These 1000 limits are combined to produce Figures \ref{CC}--\ref{slacslims}; the histograms show the distribution of upper and lower 68\% confidence limits. The medians are taken as the expected limits and the 16th and 84th percentile are taken as the errors upon the expected limits (68\% CL).

Using the \citet{bethermin} model and the velocity dispersion function of a modified Schechter function fit to the Sloan Digital Sky Survey \citep[SDSS;][]{sdss} data by \citet*{choi},
\begin{equation}
{\text{d}n}~=~ \phi_* \left({\sigma \over \sigma_*}\right)^\alpha \exp \left[-\left({{\sigma}\over{\sigma_*}}\right)^\beta \right] {\beta \over \Gamma(\alpha/\beta)}{\text{d}\sigma \over \sigma},
\label{dndsig}
\end{equation}
where $\phi_*~=~8.0 \times 10^{-3}h^3 $Mpc$^{-3}$, $\sigma_*~=~161$ kms$^{-1}$, $\alpha~=~2.32$ and $\beta~=~2.67$ for galaxy scale lenses\footnote[3]{we do not allow this function to evolve with redshift}, we calculate that typically 1 in every 250 red ellipticals will be a lens for sources with S$>$0.3~mJy at 1.1~mm. Whilst \citet{gavazzi} showed that lenses represent a population with bias towards higher mass, one can take their calculations of this bias to estimate that roughly 1\% of typical galaxy lenses will yield a second source with S$>$0.3~mJy at 1.1~mm. As such we find that a blind search of galaxy scale lenses is unlikely to yield double source plane systems unless deeper sensitivities are probed. However, by looking at surveys that preferentially select more massive systems (groups, or the rare very massive elliticals) we can increase that rate. As an example, the Cosmic Horseshoe \citep{belokurov2007} has a $\sim$30\% chance of a S$>$0.3~mJy (at 1.1~mm) object lying behind it, whilst a [$z~~=~~0.2$, $\sigma~=~350$kms$^{-1}$] elliptical will lens a S$>$0.3~mJy source $\sim$6\% of the time.

\begin{figure}
\includegraphics[width=\columnwidth]{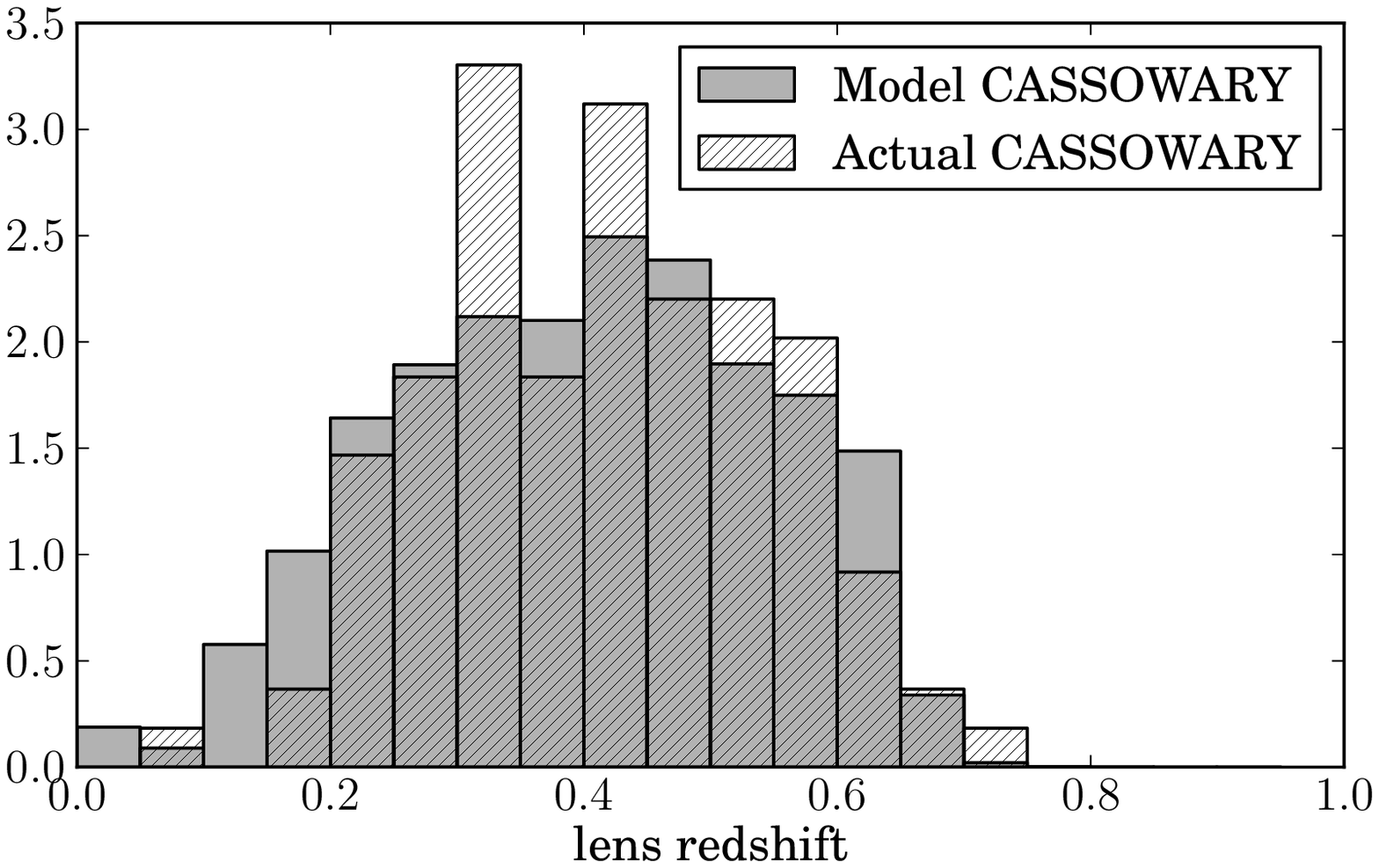}
\caption{The lens population of our {\sc CASSOWARY}-like catalogue (shown in grey), compared to the known redshifts of 106 {\sc CASSOWARY} lenses (shown hatched; not all of these are available in the currently published literature). The simple model detailed in Section \ref{cassec} broadly replicates the known CASSOWARY lens redshift distribution.}
\label{lenses} 
\end{figure}

\subsection{Follow up of a photometrically selected wide arc catalogue}

The Cambridge and Sloan survey of wide arcs in the sky ({\sc CASSOWARY}) targets wide separation blue arcs around, or multiple blue companions to, luminous red galaxies (LRGs) in the SDSS imaging catalogue. This method preferentially selects the most massive lenses, but it also often selects complicated group-scale lenses that can be difficult to model; this is the very problem that makes the double source plane method difficult for clusters \citep[but see][]{jullo}. For small groups it may be more tractable and the large, bright images that are typical in CASSOWARY provide a lot of positional information for lens models to fit and the larger lens masses that generate these wide arcs greatly increase the prospects of finding a second source. The wide arcs tend to be at relatively high redshifts, up to about $z~=~2$, although there are also some arcs with $z<1$. For the high $z_{\text{s}1}$ sources, a useful follow up strategy is to look for objects intermediate between known lens and source (probably in the optical), whilst the population of low redshift wide arcs should be followed up to find a high redshift background source (probably in the millimetre). Selecting the optimal transition point depends on the position of the lens and known source, but for simplicity we make forecasts for the constraints possible from observing the whole catalogue in the optical and in the millimetre.

\label{cassec}
\begin{figure}
\includegraphics[width=\columnwidth]{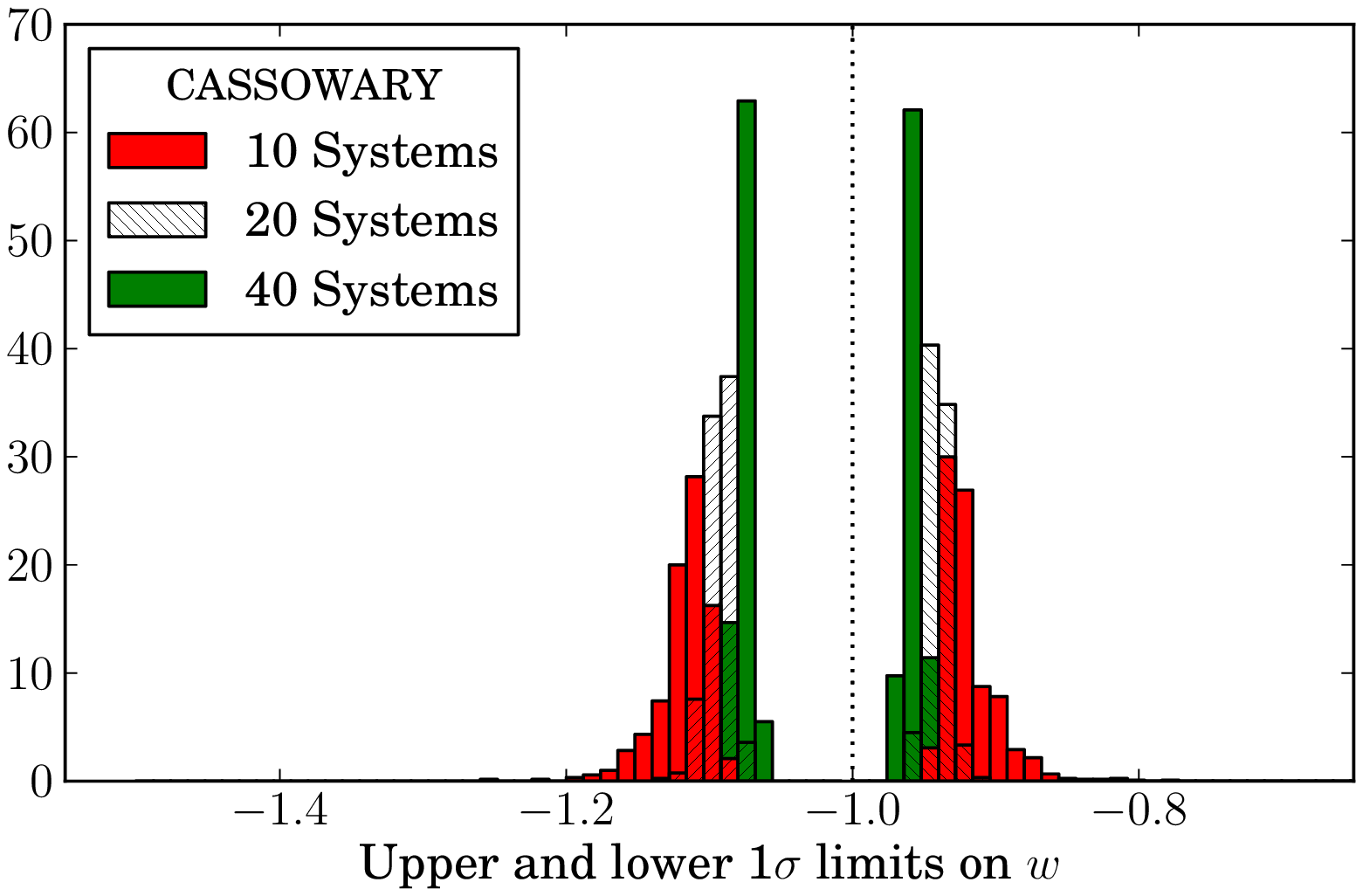}
\caption{The upper and lower bounds on the 68 percent confidence region, obtained using 1000 selections of 10 (red), 20 (hatched white) and 40 (green) double source systems. The double source systems were generated using the {\sc CASSOWARY}-like catalogue and the COSMOS photometric redshift catalogue down to $i_{\text{lim}}$~=~22. There are no green bars concealed behind the red bars.}
\label{CC} 
\end{figure}
\begin{figure}
\includegraphics[width=\columnwidth]{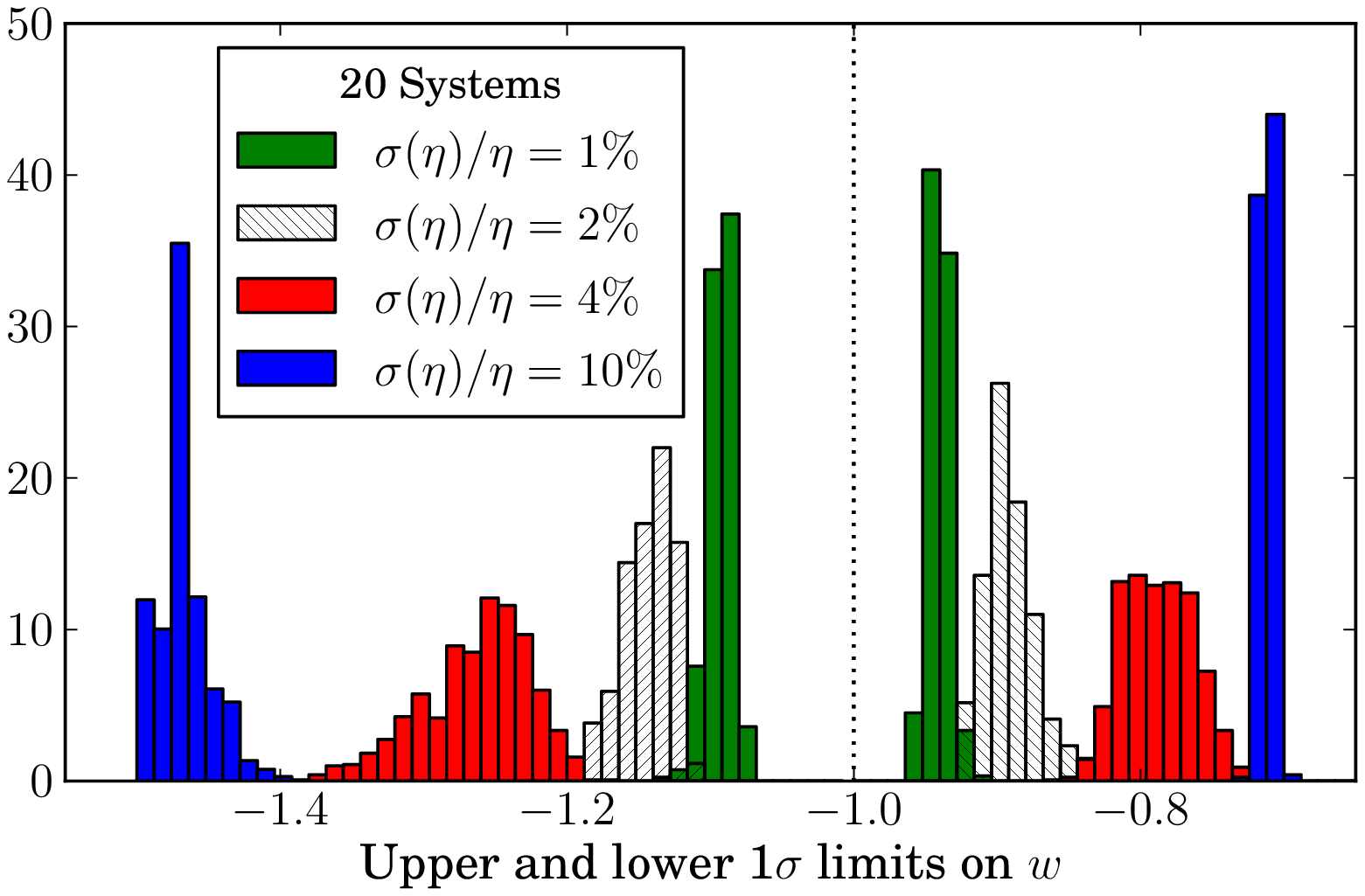}
\caption{As Figure~\ref{CC} but with 20 double source plane systems and different uncertainties on $\eta$: $\sigma(\eta)/\eta~=~1\%$ in green, $\sigma(\eta)/\eta~=~2\%$ in hatched white, $\sigma(\eta)/\eta~=~4\%$ in red, $\sigma(\eta)/\eta~=~10\%$ in blue. The $x$ scale is the same as in Figure~\ref{CC}.
}
\label{CassAcc} 
\end{figure}

Much of the {\sc CASSOWARY} deep optical imaging followup is already being conducted to study the lens and source (i.e., for non-cosmographical reasons) and this is likely to continue. Several of the lenses are known to contain two or more unique optical sources, though spectra have not always been obtained for the sources. The {\sc CASSOWARY} catalogue is still undergoing spectroscopic follow up, so there are currently few systems with known $\sigma_{\text{l}}$ and $z_{\text{s}1}$; this precludes us from estimating the {\sc CASSOWARY} selection function simply by taking the known systems [$z_{\text{l}}, z_{\text{s1}}$]. Instead, we generate a mock {\sc CASSOWARY}-like system of group-scale lenses. We assume a constant comoving density of potential lenses and take the luminosity distribution for the central galaxy as the red ellipticals luminosity function of \citet{driver}
\begin{equation}
{\dee n \over \dee \text{M}} \propto \left(10^{0.4(\text{M}-\text{M}_*)}\right)^{\alpha+1}\exp\left[-\left( 10^{0.4(\text{M}_*-\text{M})} \right)\right]
\end{equation}
where M is the absolute $B$-band magnitude, M$_*=-19.02 + 5\log_{10}(h)$ and $\alpha=-0.26$. The lensing cross section of each potential lens is then projected onto the COSMOS population in the same way as Equation \ref{integral}, except we are evaluating the probability of being a lens $P_i(L)$ in place of $P_i(D|L)$. The probability of being a {\sc CASSOWARY}-like lens requires us to make selection cuts on our model lens-source sample and a system is only included if it passes the following cuts: lens apparent magnitude, $r<$ 21.5; Einstein diameter greater than 3$''$; a limiting total arc magnitude of $r<$ 22; arc colour of $g-r <0.5$; lens absolute magnitude  M$_{\text{B}}<$ --21.5. The last cut is made because typical LRG selections exclude low luminosity objects, as is seen clearly in the luminosity function of \citet{eisenstein2001}. Once this procedure has been applied, we are able to create a weighted population of $\sigma_{\text{l}}$ and $z_{\text{s}1}$ for each $z_{\text{l}}$. Our catalogue broadly replicates the known lens redshift of 106 {\sc CASSOWARY} lenses (and lens candidates), as shown in Figure~\ref{lenses}. Since the purpose of generating this population is to estimate future cosmological constraints rather than to perfectly replicate any particular current survey, we are unconcerned by the simplicity of this {\sc CASSOWARY}-like model. With this mock catalogue of [$z_{\text{l}}$, $\sigma_{\text{l}}$] we can generate second sources using the method outlined above, the weights for each double source plane system are now given by $P_i(L) \times P_i(D|L)$ if the lens and first source pass all the selection cuts, and is zero otherwise.

\begin{table}
\caption{Expected 1$\sigma$ upper and lower bounds on $w$, given $n$ {\sc CASSOWARY} double lens systems. The lens and first sources are generated using the CASSOWARY--like model and the second sources come from the deep optical catalogue ($i<$ 22), and the \citet{bethermin} millimetre model (S $>$ 0.3~mJy). The Section marked ``Combination" has second sources taken from both the second source populations, with a weighting of approximately three optical second sources per millimetre second source.}
\label{castable}
\begin{tabular}{@{}lccc}
\hline
\multicolumn{4}{l}{Optical, $\sigma(\eta)/\eta~=~1\%$}\\
\hline
$n$& 10&20&40\\
\hline
$w_{\text{upper}}$&$-0.902_{-0.018}^{+0.019}$&$-0.926_{-0.004}^{+0.006}$&$-0.936_{-0.003}^{+0.003}$
\vspace{2 mm}\\
$w_{\text{lower}}$&$-1.093_{-0.017}^{+0.014}$&$-1.066_{-0.014}^{+0.006}$&$-1.050_{-0.005}^{+0.007}$
\\
\hline
\multicolumn{4}{l}{Optical, $\sigma(\eta)/\eta \neq 1\%$}\\
\hline
$n$&20&20&20\\
$\sigma(\eta)/\eta$&2\%&4\%&10\%\\
\hline
$w_{\text{upper}}$&$-0.871_{-0.013}^{+0.023}$&$-0.765_{-0.024}^{+0.018}$&$-0.705_{-0.005}^{+0.003}$
\vspace{2 mm}\\
$w_{\text{lower}}$&$-1.121_{-0.018}^{+0.013}$&$-1.238_{-0.040}^{+0.033}$&$-1.469_{-0.020}^{+0.028}$
\\
\hline
\hline
\multicolumn{4}{l}{Millimetre, $\sigma(\eta)/\eta~=~1\%$}\\
\hline
$n$&5&7&15\\
\hline
$w_{\text{upper}}$&$-0.868_{-0.027}^{+0.054}$&$-0.886_{-0.022}^{+0.037}$&$-0.921_{-0.005}^{+0.018}$
\vspace{2 mm}\\
$w_{\text{lower}}$&$-1.125_{-0.053}^{+0.023}$&$-1.107_{-0.032}^{+0.019}$&$-1.079_{-0.014}^{+0.013}$
\\
\hline
\hline
\multicolumn{4}{l}{Combination, $\sigma(\eta)/\eta~=~1\%$}\\
\hline
$n$&15&30&50\\
\hline
$w_{\text{upper}}$&$-0.921_{-0.005}^{+0.016}$&$-0.932_{-0.003}^{+0.003}$&$-0.939_{-0.001}^{+0.002}$
\vspace{2 mm}\\
$w_{\text{lower}}$&$-1.079_{-0.014}^{+0.013}$&$-1.057_{-0.006}^{+0.006}$&$-1.043_{-0.006}^{+0.002}$
\\
\hline
\label{cassresult}
\end{tabular}
\end{table}

If the lenses are assumed to have an effective velocity dispersion of $\langle \sigma ^4 \rangle ^{1 \over 4} \approx 350$ kms$^{-1}$, we find 7\% to also be lensing an S$>$0.3~mJy millimetre source. For the {\sc CASSOWARY} objects already observed, 3 of 15 have so far been discovered to have a second arc. We assume therefore that 7 millimetre sources and 20 optical sources would be yielded by a fullscale follow-up campaign. We make forecasts for the cosmographic constraints possible with 5, 7, and 15 millimetre sources and with 10, 20, and 40 optical double source systems. Figure~\ref{CC} shows the likelihood distribution of the upper and lower limit on $w$ with 10, 20, and 40 optical second sources ($\sigma(\eta)/\eta~=~1\%$). \comments{and Figure~\ref{CB7} shows the same but with 7 millimetre sources (S$>$0.3~mm).} Finally we combine the datasets (still assuming $\sigma(\eta)/\eta~=~1\%$) to draw 15, 30, and 50 double source systems. Because of the complexity of group-scale structures it may prove impossible to measure $\eta$ to 1\% uncertainty; Figure~\ref{CassAcc} shows the likelihood distributions for 20 double optical sources if a 1\%, 2\%, 4\% or 10\% uncertainty on $\eta$ can be achieved; the results are similar to Figure~\ref{unc} in that the constraints on $w$ depend significantly upon $\sigma(\eta)$, but given 20 systems even 4\% uncertainties on $\eta$ can provide meaningful constraints on $w$. If $\sigma(\eta)/\eta$ is as high as $10\%$ a population of many more than 20 double source plane systems would be needed to make an important contribution to cosmography. All of the results for this {\sc CASSOWARY}-like model are shown in Table \ref{castable}.
\comments{

\begin{figure}
\includegraphics[width=\columnwidth]{CB7.eps}
\caption{As Figure~\ref{CC}, but with 7 double source plane systems, and the second source generated using the B\'ethermin model for 1.1~mm continuum observations down to a depth of 0.3~mJy. The forecast upper limit is $-0.914_{-0.012}^{+0.0049}$ and the forecast lower limit is $-1.124_{-0.041}^{+0.017}$.}
\label{CB7}
\end{figure}

}
\subsection{Follow up of a SLACS type catalogue}
\begin{figure}
\includegraphics[width=\columnwidth]{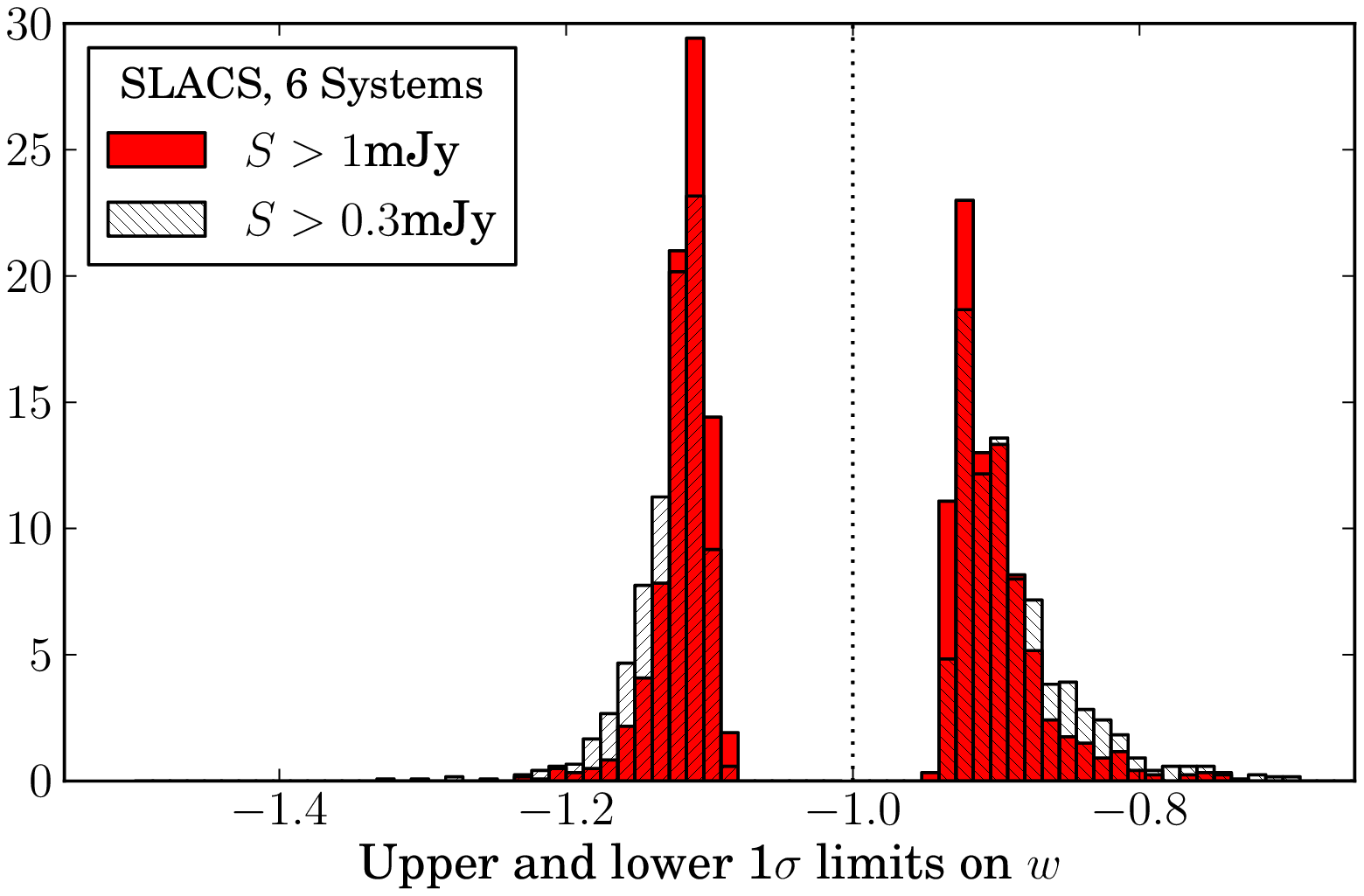}
\caption{The upper and lower bounds on the 68 percent confidence region, obtained using 1000 selections of six double source systems. The double source systems were generated using the SLACS catalogue and the B\'ethermin et al. model for 1.1~mm continuum observations the grey bars are for sources selected at S$>$0.3~mJy and the red bars are from sources selected at S$>$1~mJy. The constraining power degrades by $\sim$1 percentage point, for S$>$0.3~mJy compared to the S$>$1~mJy results.}
\label{slacslims} 
\end{figure}

\begin{figure}
\includegraphics[width=\columnwidth]{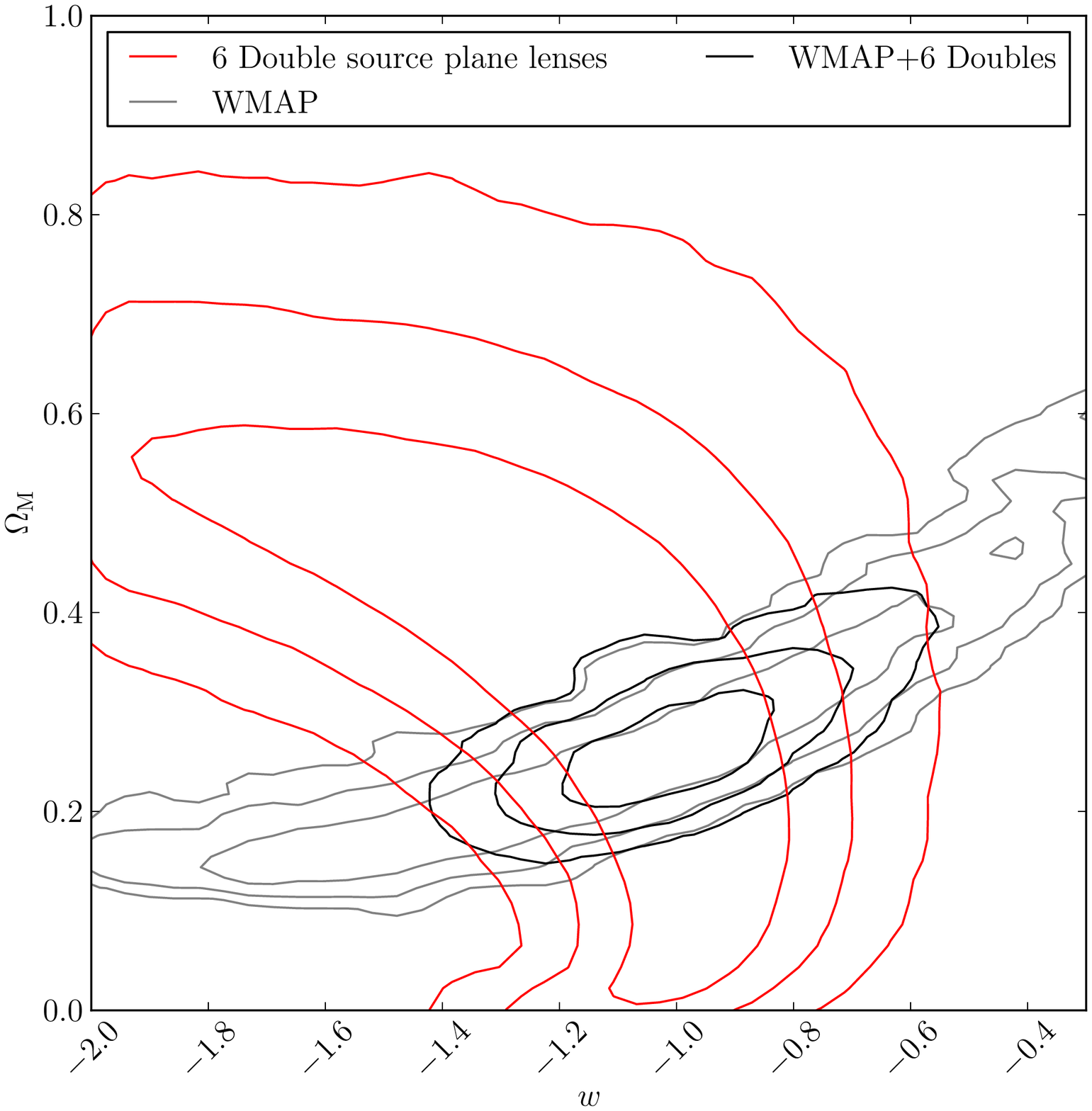}
\caption{The $w$--$\Omega_{\text{M}}$ plane for a typical selection of 6 SLACS--type double source plane systems ({\it red}), with WMAP ({\it grey}) and the combination of both ({\it black}). The six systems are listed in Table \ref{systems}. The confidence regions have shrunk significantly compared to one system (Figure~\ref{sim}) and are still orthogonal to the WMAP constraints.}
\label{6slacs} 
\end{figure}

The Sloan Lens ACS (SLACS) Survey selected strong lensing candidates using the SDSS spectroscopic survey. Specifically targeting bright early type galaxies, SLACS takes advantage of the $3''$ diameter spectroscopic fibres used by the SDSS and looks for emission line features that suggest a higher redshift component than the target galaxy \citep{bolton2004, bolton2006, bolton2008, auger2009}; these targets are then followed up using ACS and WFCP2 on the {Hubble Space Telescope (HST)}. The selection function of SLACS is complicated \citep{dobler}, but large $\Delta z_{\text{l-s1}}$ systems are selected against since the emission lines are often redshifted out of the bandpass, and the finite diameter of the fibre requires small $\Delta z_{\text{l-s1}}$ for the most massive lenses. As Section \ref{onesys} showed, bias towards low $\Delta z_{\text{l-s1}}$ is desirable for finding optimal double source systems. The selection of lower mass objects (compared to group-scale lenses, for example) decreases the prospect of finding a second source behind the SLACS system because the lensing cross section scales with mass squared; however the ideal follow up candidates (small $\Delta z_{\text{l-s}}$, high mass) still show up in the SLACS sample--because of their small Einstein radius--despite their high mass. Using the \citet{bethermin} model, a full millimetre follow up of the current SLACS catalogue down to 1~mJy should produce $\sim$$1.5$ new double source systems, whilst observing to 0.3~mJy would yield $\sim$$3.0$. Since the catalogue is growing, and this is a conservative flux limit given the observing capability of ALMA, we forecast the expected constraints on $w$ given by 1, 2, and 6 double source systems found with S$>$1~mJy and 3, 6, and 12 systems found with S$>$0.3~mJy. The results are summarised in Table \ref{slacsresult}. Figure~\ref{slacslims} shows the likelihood distribution of the upper and lower limits on $w$ given six double source systems with S$>$0.3~mJy and S$>$1~mJy. Figure~\ref{6slacs} shows the $w$--$\Omega_{\text{M}}$ plane for a typical selection of 6 SLACS-type S$>$0.3~mJy double sources; the six ``typical" systems are listed in Table \ref{systems}.
\begin{table}
\caption{Expected 1$\sigma$ uncertainty on $w$, given $n$ SLACS double lens systems, detected at a flux limit of S. In all cases $\sigma(\eta)/\eta~=~1\%$. }
\label{symbols}
\begin{tabular}{@{}lccc}
\hline
\multicolumn{4}{l}{S~=~1~mJy}\\
$n$& 1&2&6\\
\hline
$w_{\text{upper}}$&$-0.720_{-0.109}^{+0.030}$&$-0.783_{-0.076}^{+0.080}$&$-0.891_{-0.018}^{+0.028}$
\vspace{2 mm}\\
$w_{\text{lower}}$&$-1.237_{-0.236}^{+0.092}$&$-1.172_{-0.101}^{+0.048}$&$-1.098_{-0.017}^{+0.015}$
\\
\hline
\hline
\multicolumn{4}{l}{S~=~0.3~mJy}\\
$n$&3&6&12 \\
\hline
$w_{\text{upper}}$&$-0.822_{-0.049}^{+0.100}$&$-0.880_{-0.022}^{+0.044}$&$-0.917_{-0.009}^{+0.017}$
\vspace{2 mm}\\
$w_{\text{lower}}$&$-1.142_{-0.072}^{+0.032}$&$-1.106_{-0.022}^{+0.018}$&$-1.078_{-0.011}^{+0.013}$
\\
\hline
\label{slacsresult}
\end{tabular}
\end{table}

\begin{table}
\caption{The typical set of 6 SLACS--type lenses. These are the 6 systems that gave the median constraints on $w$ for WMAP+6 double source plane lenses in Table \ref{slacsresult}. One SLACS system has reappeared ($z_{\text{l}}~=~0.227$, $z_{\text{s}}~=~0.931$), but such repeats are not excluded by the weighted selection of lens-source systems; neither Systems 2 or 4 are likely to play a significant part in constraining $w$ due to their large $\Delta z_{\text{l-s1}}$. For cosmography, System 5 is the best system by some margin, although systems 1 and 3 are also significant when constraining evolving models.}
\begin{tabular}{@{}lcccccc}
\hline
System&1&2&3&4&5&6\\
z$_{\text{l}}$&0.440&0.227&0.195&0.227&0.194&0.111\\
z$_{\text{s1}}$&1.192&0.783&0.632&0.931&0.446&0.316\\
z$_{\text{s2}}$&3.859&0.931&2.724&1.766&2.058&0.463\\
\hline
\end{tabular}
\label{systems}
\end{table}

Directly comparing the two instances of six double source lenses shows some degrading of the constraint on $w$ as the flux limit is lowered. This is due to the large number of low flux, low redshift objects predicted by \citet{bethermin} which are individually less useful for cosmography than the higher-redshift systems. A shallow survey of lots of lenses would select fewer of these low redshift millimetre sources (and thus be optimal for cosmography), but since there are very few known (simple) high mass lenses, such a survey is currently impossible. Despite the fact that SLACS provides much better cosmography per double source plane system than the {\sc CASSOWARY}-type, our results are dominated by the few SLACS objects with very high velocity dispersion and close first source: the lower velocity dispersion objects are unlikely to be double source plane systems and there are no high velocity dispersion systems with intermediate or large lens--source separation. Indeed, many high mass lenses will be missed by SLACS due to the source arcs falling outside the spectroscopic fibre. These high mass systems will be crucial if a large number of double source systems are to be discovered efficiently, so follow up observations of all the (simply-modeled) high mass systems in strong lens catalogues should be conducted.

\section{Beyond $w$CDM}\label{complications}

\begin{figure}
\includegraphics[width=\columnwidth]{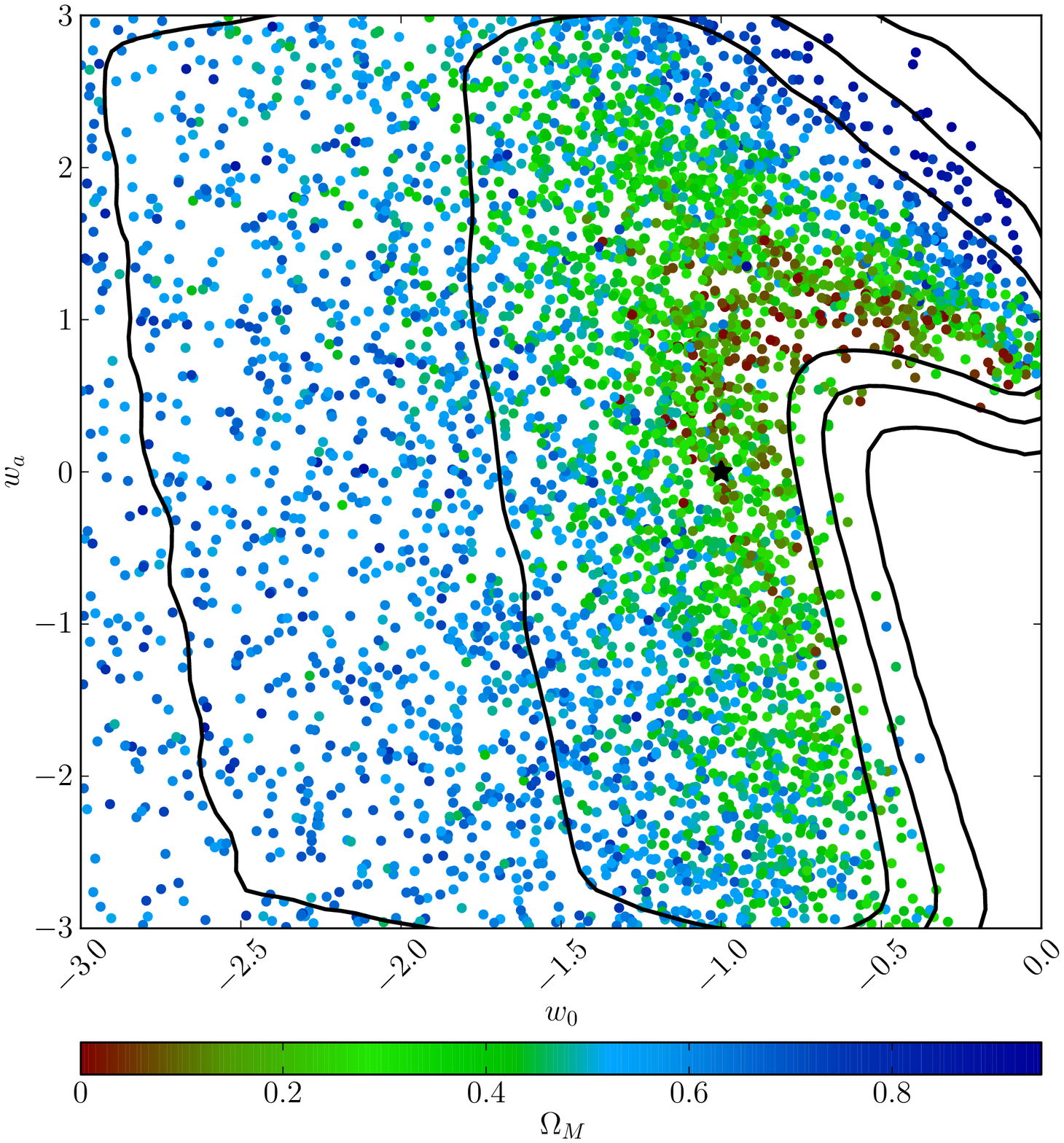}
\caption{Forecast parameter constraints for the $w(z)~=~w_0+w_a(1-a)$ model. Constraints are from the six systems listed in Table \ref{systems} assuming $\sigma(\eta)/\eta~=~1\%$, and uniform priors [$0 \le \Omega_{\text{M}} \le 1$; $-3 \le w_0 \le {0}$; $-3 \le w_a \le +3$]. The plot shows 1, 2 and 3 $\sigma$ regions in the $w_0-w_a$ plane, marginalised over $\Omega_{\text{M}}$, the samples are coloured by $\Omega_{\text{M}}$; a stronger prior on $\Omega_{\text{M}}$ is needed to improve the constraint on $w_0$ and $w_a$.}
\label{waconf} 
\end{figure}

Up to this point we have only investigated models with constant equation of state. \citet{caldwell} first suggested the possibility of an evolving equation of state in the context of a quintessence model. We now investigate the power of double source plane lenses to constrain $w(z)$ models. We adopt the parametrisation
\begin{equation}
\begin{split}
w(z)&~=~w_0+w_a(1-a)\\
&~=~w_0+w_a\left({z \over {1+z}}\right), 
\end{split}
\label{wa}
\end{equation}
as introduced in \new{\citet{CP} and} \citet{linder2003}. Our fiducial cosmology is a special case of this parametrisation with $w_0~=~-1$ and $w_a~=~0$. This parametrisation allows us to solve the integral in Equation \ref{Iz} analytically:
\begin{equation}
e^{I(z)}~=~(1+z)^{3(1+w_0+w_a)}\exp\left({-3 w_a z\over 1+z}\right).
\end{equation}
Taking the 6 systems in Table \ref{systems}, $\sigma(\eta)/\eta~=~1\%$, and uniform priors [$0 \le \Omega_{\text{M}} \le 1$; $-3 \le w_0 \le -{1 \over 3}$; $-3 \le w_a \le +3 $], our MCMC forecast finds the confidence intervals shown in Figure~\ref{waconf}. In high $\Omega_{\text{M}}$ universes, the dark energy equation of state is poorly constrained by measuring $\eta$. Marginalising over all $\Omega_{\text{M}}$ leaves very weak constraints in the  $w_0$--$w_a$ plane because of the high  $\Omega_{\text{M}}$ samples. With a prior on $\Omega_{\text{M}} < 0.5$ the $w_0$--$w_a$ plane shows a boomerang shaped degeneracy, which arises from taking the ratio of angular diameter distances. A universe with $w_0+w_a > 0$ has no era of matter domination and is strongly excluded by CMB constraints; coincidentally this is the region for double source plane constraints that is  allowed by taking angular diameter distance ratios. The six double source plane systems alone give $w_0~=~-1.177^{+0.460}_{-0.808}$, $w_a~=~0.485^{+1.455}_{-2.353}$. Combined with a CMB prior -- forecast for Planck -- the 1 $\sigma$ constraints are $w_{0}$ = $-0.802^{+0.495}_{-0.590}$, $w_a$ = $-0.655^{+1.228}_{-1.175}$.

\begin{figure*}
\includegraphics[width=\textwidth]{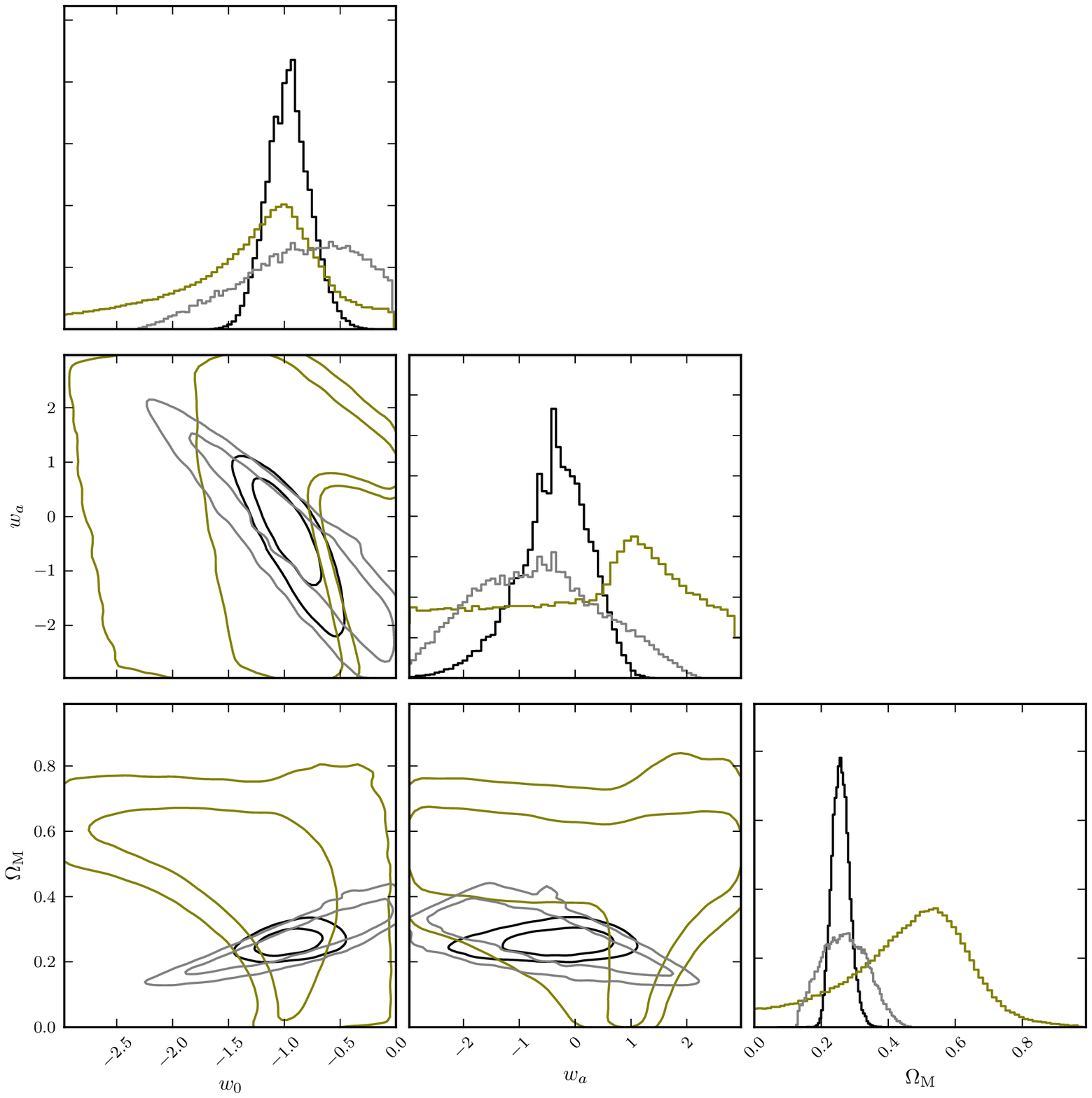}
\caption{Forecast cornerplot for the $w(z)~=~w_0+w_a(1-a)$ model. On diagonal subplots shows the marginalised likelihood for each parameter. Off-diagonal subplots show the two-dimensional 1 and 2 $\sigma$ confidence contours, marginalised over the third parameter. The grey contours are forecast for Planck (CMB lensing and ISW effects have been included). Olive shows the constraints from 6 double source plane systems in Table \ref{systems} assuming $\sigma(\eta)/\eta~=~1\%$. Black contours are forecast for the combination of Planck with the 6 double source plane systems.\comments{ Note that since the allowed parameter space is complicated, marginalisation does not always recover the fiducial cosmology.}}
\label{planckwa} 
\end{figure*}
\begin{figure*}
  \begin{center}
      \resizebox{70mm}{!}{\includegraphics{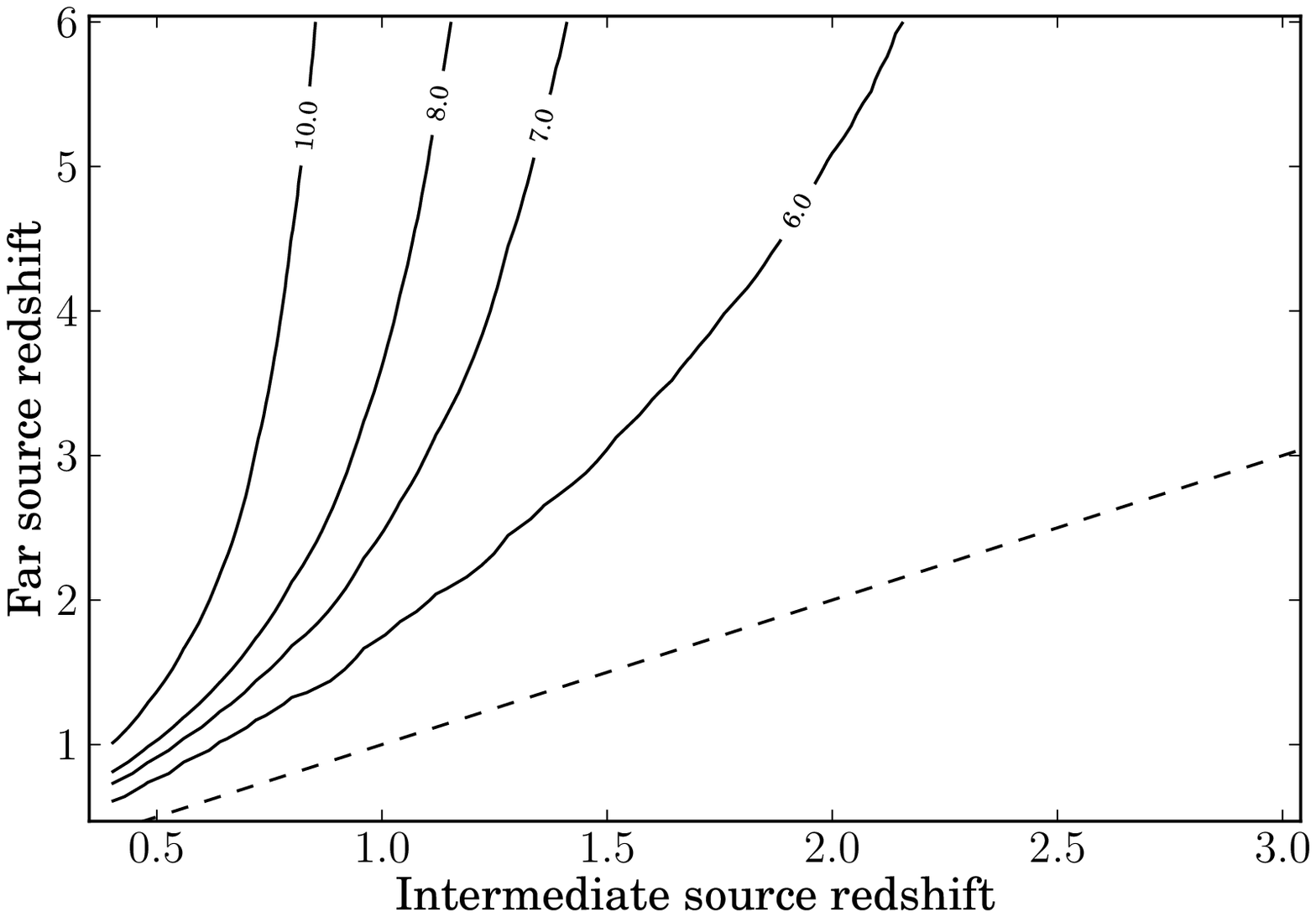}}
      \hspace{10 mm}
      \resizebox{70mm}{!}{\includegraphics{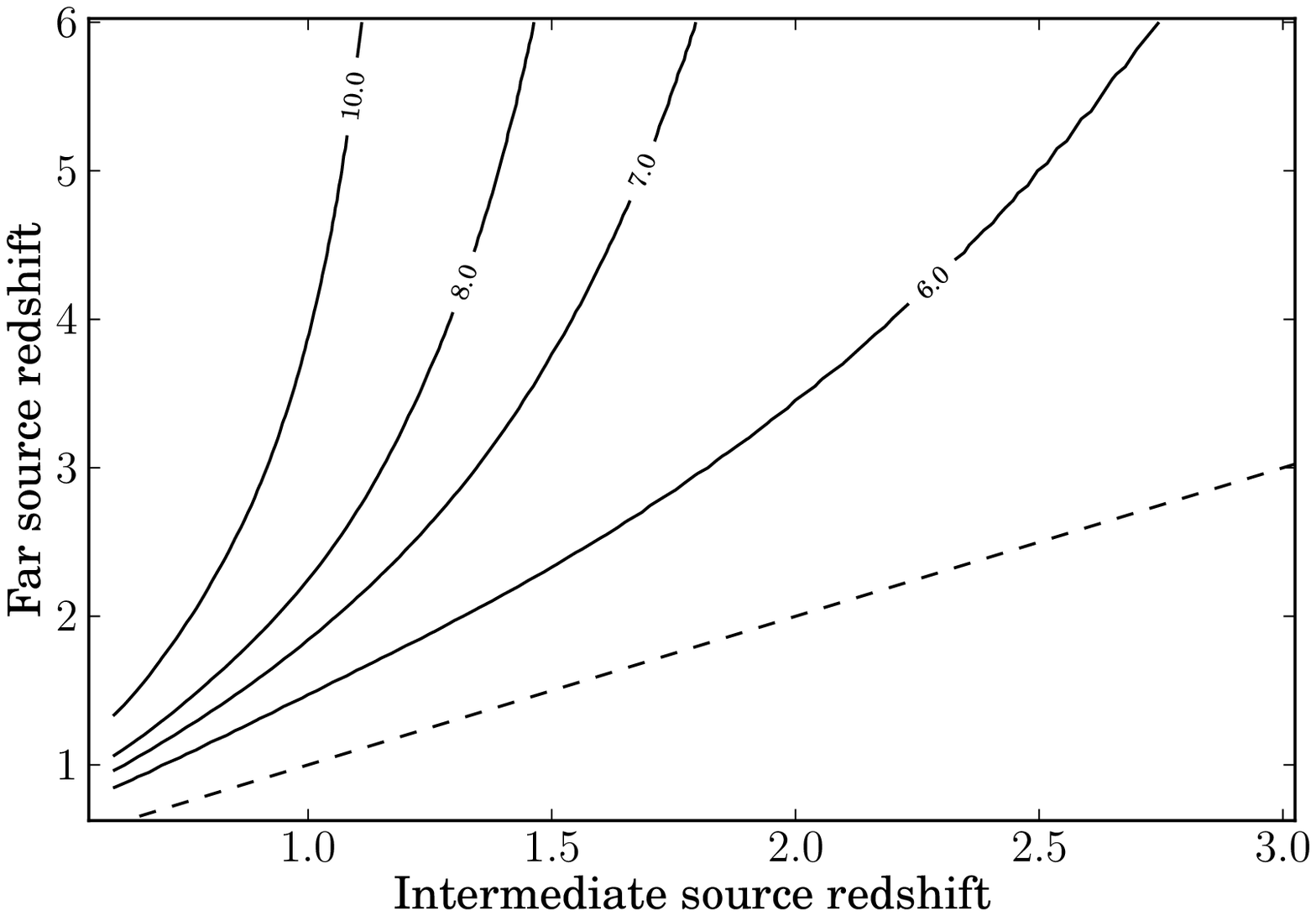}}
  \end{center}
\caption{Figure of merit contour plots for the  $w(z)~=~w_0+w_a(1-a)$ model, with FOM$~=~{6.17\pi / A_{95}}$. Constraints are for a single double source plane lens assuming $\sigma(\eta)/\eta~=~1\%$, combined with a forecast Planck dataset. {\it Left} Lens redshift $z_{\text{l}}$~=~0.35. {\it Right} Lens redshift $z_{\text{l}}$~=~0.55. A large second source redshift and small separation between lens and first source gives the best FOM.}
\label{FOMintuition} 
\end{figure*}

For the Planck MCMC forecasts, we use the seven frequency channels that have polarisation sensitivity, with noise levels given in the Planck Bluebook \citep{bluebook}. In addition we include a reconstruction of the CMB weak lensing potential, with statistical noise calculated using the optimal quadratic estimator of \citet{okamoto}. We use information from the spectra up to a maximum angular multipole of 2500, and assume that 65\% of the sky is observed. We sampled the likelihood using the publicly-available CosmoMC package \citep{lewis}, with fiducial spectra calculated with CAMB \citep{camb}. \new{Evolving dark energy models were implemented with the PPF CAMB module of \citet{fang}}. In Figure~\ref{planckwa} we show the Planck and 6 double source plane system constraints for each parameter pair in the flat $w_a$ model, marginalized over the third parameter. 

It is clear from Figures \ref{waconf} and \ref{planckwa} that $w_0$ and $w_a$ are degenerate; as such we use the figure of merit (FOM) as defined in Equation~6 of \citet[][]{mortonson} to quantify the constraining power:
\begin{equation}
\mathrm{FOM}~=~{6.17\pi \over A_{95}}, 
\end{equation}
where $A_{95}$ is the area enclosed by the 95\% confidence interval in the $w_0$--$w_a$ plane marginalised over all other parameters. Using this definition the FOM for Planck plus the 6 lenses in Table \ref{systems} is 14.2. Using WMAP+6 systems lowers the figure of merit to 7.6. When asserting spatial flatness, \citet[][]{mortonson} find FOM~=~15 for WMAP plus the Union supernovae compilation \citep{union}, combined with the baryon acoustic oscillation data of \cite{eisenstein} and an $H_0$ prior. 

Figure~\ref{FOMintuition} shows the FOM contour plot for Planck plus one double source plane system with $\sigma(\eta)/\eta~=~1\%$; we find that for $w_0, w_a$ models the figure of merit increases as $z_{\text{s2}}$ increases and as $\Delta z_{\text{l-s1}}$ decreases. We also find that the $w_0 - w_a$ degeneracy increases (away from vertical in Figure~\ref{planckwa}) with increasing lens redshift, conversely below $z_{\text{l}} \approx 0.2$ the degeneracy is almost broken, but the FOM rapidly degrades as $z_{\text{l}}$ decreases. Since the degeneracy between parameters changes with the redshift of lens and sources, the FOM is likely to grow quickly with sample size. \new{To strongly constrain evolving models it will be important to have a good coverage of lens redshifts. The ongoing BELLS survey \citep{BELLS} uses very similar selection methods to SLACS, but targets higher redshift potential lenses; their 25 definite and 11 probable galaxy-galaxy systems have lens redshifts 0.4 $\lesssim$ z $\lesssim$ 0.7. These higher redshift systems should be especially useful for incressing the double source plane FOM -- if a second source can be found behind any of them.}

Throughout this work we have asserted a spatially flat universe, but allowing $\Omega_k$ to be free weakens the constraints on $w$. Without asserting spatial flatness we find that WMAP plus the 6 typical systems give the following results for a constant equation of state: $w~=~-1.11_{-0.33}^{+0.17}$. For non-flat evolving equation of state models our FOM for Planck plus 6 lenses degrades slightly to 12.3. Double source plane systems place no significant additional constraints on $\Omega_k$ once the Planck results are included due to the excellent precision with which CMB lensing constrains flatness.
\begin{figure}
  \begin{center}
    \begin{tabular}{c}
      \resizebox{\columnwidth}{!}{\includegraphics{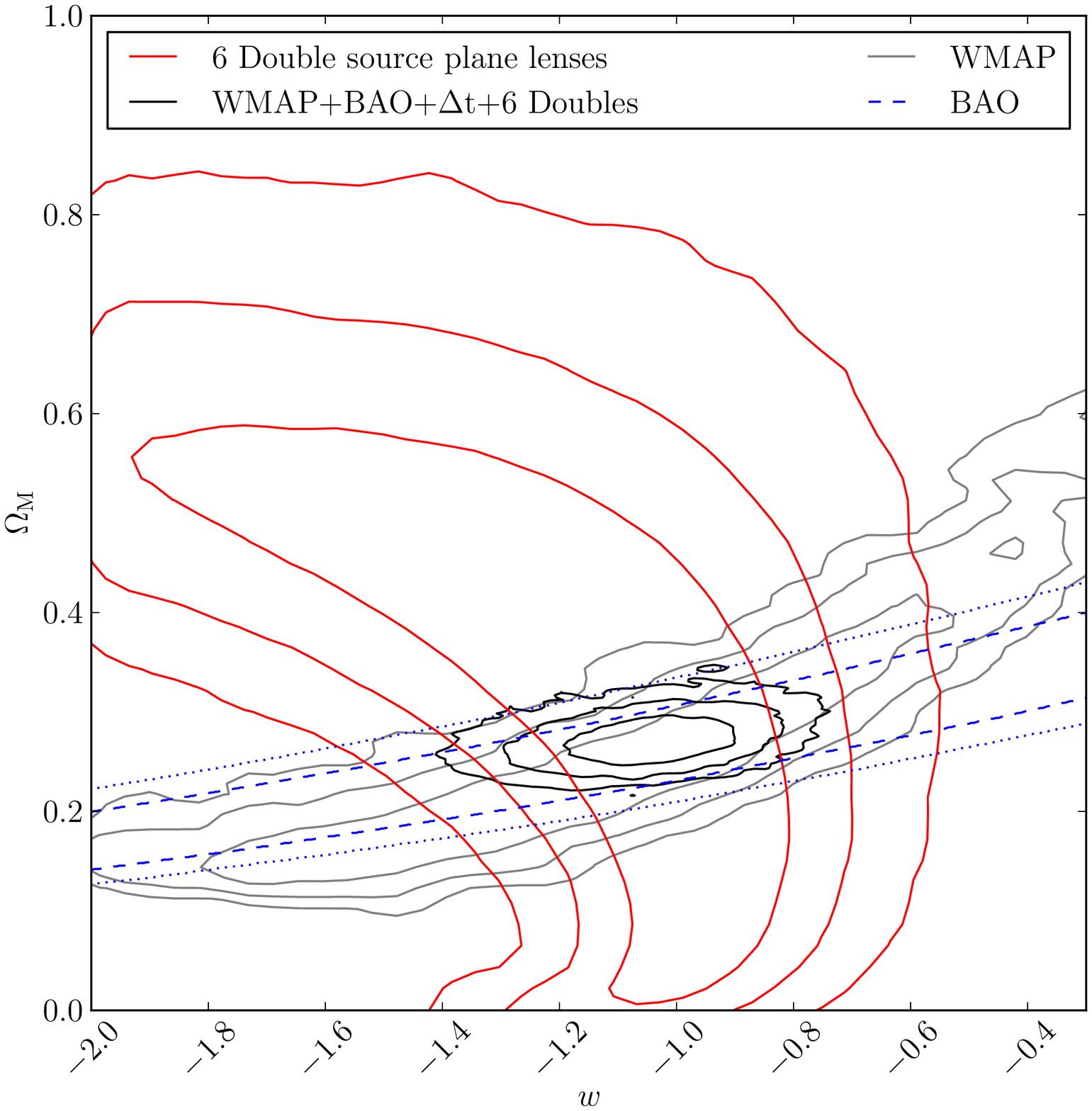}} \\
      \resizebox{\columnwidth}{1.70 in}{\includegraphics{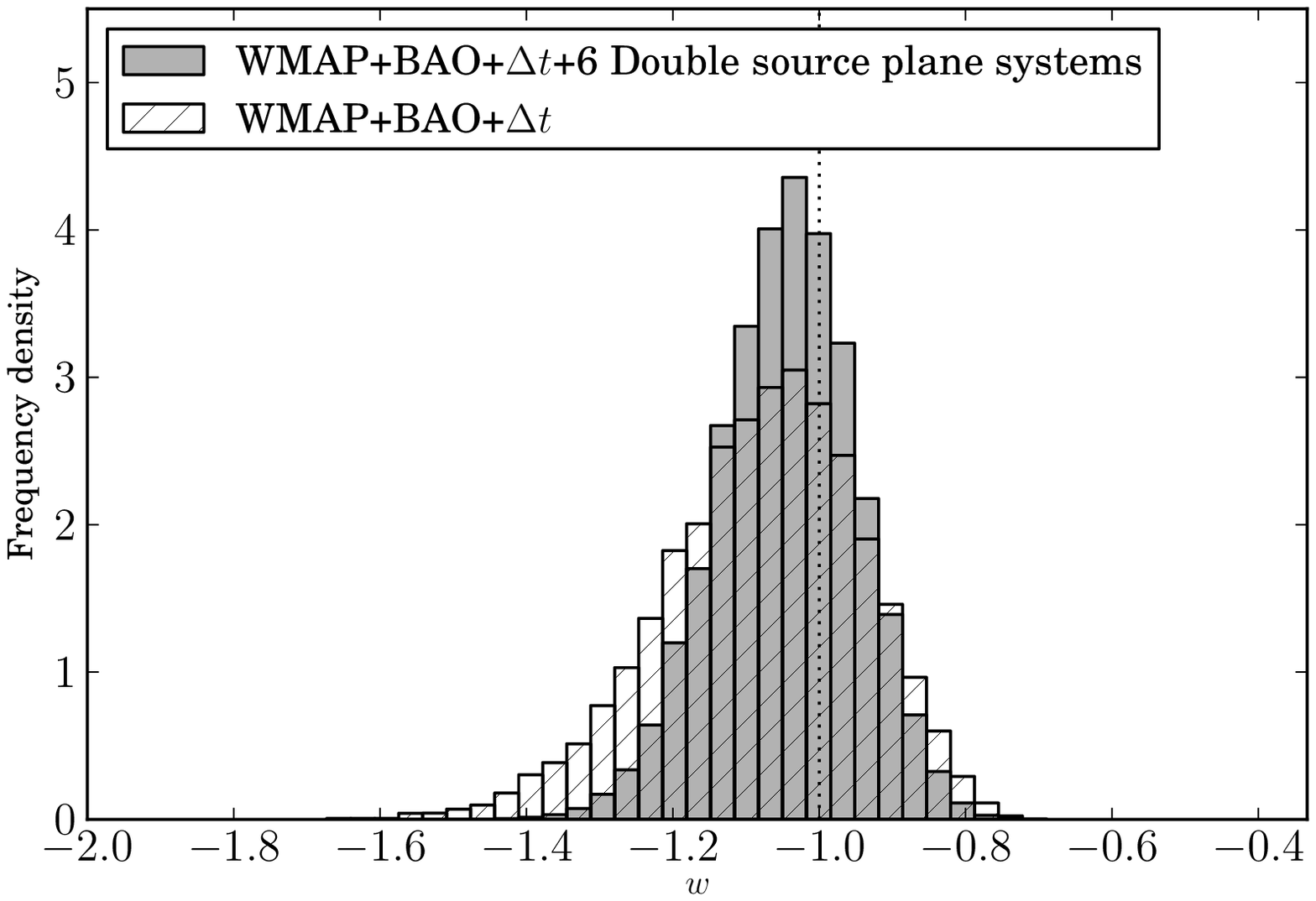}} \\
    \end{tabular}
  \end{center}
\caption{{\it Top} The $w$--$\Omega_{\text{M}}$ plane for a typical selection of 6 SLACS--type double source plane systems ({red}), with WMAP ({grey}) and the BAO constraints of \citet{eisenstein} (dashed blue). The combination of WMAP+BAO+$\Delta t$ and the 6 SLACS lenses is shown in {black}. The \citet{suyu} results are not shown in this figure, but are included in the black contour. {\it Bottom} The marginalised distribution of $w$ for WMAP+BAO+$\Delta t$ (hatched) and the improvement gained when adding 6 SLACS--type double source plane systems (grey).}
\label{multiple}
\end{figure}
\section{Discussion and conclusion}
Strongly lensed double source systems and the CMB are not the only datasets to constrain $w$; Supernovae (SNe), Baryon Acoustic Oscillations (BAO), galaxy clusters, and time delays in strongly lensed time variable sources ($\Delta t$) also provide constraints on $w$. Whilst only the strongly lensed double source method is independent of the Hubble constant, WMAP+BAO+$\Delta t$ can still give strong constraints on $w$ when combined with a prior on $h$ obtained using local supernovae (we have not included high redshift supernovae or galaxy cluster constraints). \citet{komatsu} used this combination of methods, finding $w~=~-1.08\pm0.13 $. In the future the constraints from strongly lensed double source plane systems can easily be combined with other datasets to give an improved constraint on $w$. The main advantages of this are the breaking of degeneracies and, perhaps more importantly, the investigation of systematics inherently afforded by using different measurements. Figure~\ref{multiple}, shows the $w$--$\Omega_{\text{M}}$ plane for a typical set of 6 SLACS type double sources (listed in Table \ref{systems}) given a uniform flat prior on $w$ and $\Omega_{\text{M}}$ and a forecast combination of WMAP+BAO+$\Delta t$+6 double source plane lenses. In this case we recover a constraint of $w~=~-1.04_{-0.09}^{+0.10}$, for our fiducial cosmology of $w~=~-1$. Setting the fiducial cosmology to $w~=~-1.08$ yields $w~=~-1.08_{-0.09}^{+0.10}$. These represent a 30\% improvement on the \citep{komatsu} result, although the $\Delta t$ and BAO constraints will improve significantly as their datasets grows and if $h$ is better constrained; this method might be slightly less competitive by the time 6 double source plane systems have been found, but a significant strength of multiple source plane lenses is as a constraint on $w$ independent of any measurement of $h$.

Most $w_0, w_a$ forecasts have used a Fisher formalism to estimate future parameter constraints \citep[e.g][]{coe, shapiro}. Whilst the Fisher forecasts are not directly comparable to our MCMC forecasts it seems that the $w_0, w_a$ degeneracy for double source plane lenses will be in a similar direction to BAO, SNe and weak lensing. For all these probes low $w_0$ correlates with positive $w_a$. However, for very low lens redshifts ($z_{\text{l}} \approx 0.2$), the double source plane constraints on $w_0$ become almost independent of $w_a$, so there is scope for complementarity, but at the expense of the constraining power of each lens. 

The problem of finding second sources will not be trivial, and the probability of finding a second source scales with mass squared. As such a dedicated search strategy should look to find second sources behind high mass lenses. Currently there are a handful of good candidate systems, and a few double source systems should be uncovered with a moderate investment of telescope time. With a larger population of known lenses, the number of very good candidates for such a search will increase, making searches for compound lenses more efficient and more useful; not all double source systems will be particularly useful for cosmography, mostly due to having both sources far beyond the lens, but a selective search strategy can minimise the time spent on these objects. One wants to draw the second source from a different population to the first to make $z_{\text{s1}}$ $\approx$ $z_{\text{s2}}$ less likely; this can either be achieved by looking in a different wavelength or to significantly greater depth. It may also not be possible to measure $\eta$ to 1\% for all objects, especially after accounting for lensing by the second source. However, the {\sc CASSOWARY} results for 20 lenses with a 4\% uncertainty on $\eta$ show that given a large sample, $w$ can still be constrained with double source plane lens systems. Further work will evaluate the systematics of measuring $\eta$, paying particular attention to the mass model of the lens and lensing by the intermediate source or other perturbers along the line of sight.

\begin{enumerate}
\item {\it Can double source plane lenses be used for cosmography?}

Double source plane lenses can be used for cosmography. We have shown that -- uniquely amongst cosmological probes -- it is possible to constrain the equation of state with double source plane lenses, independent of the Hubble parameter.\\

\item {\it What is an optimal configuration of lens and source redshifts for cosmography?}

By resampling the WMAP team's MCMC chains we have shown that, for cosmography, the key features are a low separation lens--source pair and a high redshift background source. We found cosmography to be broadly insensitive to lens redshift, so long as it is greater than about 0.2.\\

\item {\it How well could cosmography be constrained with a realistic population of double source plane lenses?}

Given our knowledge of what represents a good system for cosmology, we postulate a search strategy to look for double source plane systems; ultimately such a follow up strategy will depend on the single source plane systems (which will likely increase in number), but using the population model of \citet{bethermin} and the photometric redshift catalogue of COSMOS, our model allowed us to make forecasts for a population of double source plane systems discovered in a variety of different ways. From this population we have learnt that -- at $\sigma(\eta)/\eta~=~1\%$ -- only a handful of double source plane systems are needed to be competitive with current constraints on $w$ and that 6 systems can provide meaningful constraints for $w_0, w_a$ models. The $\Omega_{\text{M}}$--$w$ degeneracy of this method is almost orthogonal to CMB and BAO datasets, making the double source plane method complimentary to existing work.
\end{enumerate}

\section*{Acknowledgments}
We thank Michael Mortonson for advice on producing $w_0, w_a$ chains for Planck, and providing us with WMAP chains. \new{We are grateful to Sherry Suyu and the referee for suggested alterations to the original manuscript.} TEC is supported by an STFC studentship. VB acknowledges financial support from the Royal Society. AH is supported by an Isaac Newton Studentship and the Isle of Man Government.

\end{document}